\documentclass[conference]{IEEEtran}
\usepackage{graphicx} 
\usepackage{tabularx}
\usepackage{booktabs}
\usepackage{hyperref}
\usepackage{url}
\usepackage{subcaption}
\usepackage{xcolor}
\usepackage{paralist}
\usepackage{colortbl}
\begin{document}
\title{{\underline{DuoLungo:}} Usability Study of Duo 2FA}
\author{
{\rm Renascence Tarafder Prapty}\\
    University of California Irvine\\
    rprapty@uci.edu
    \and
    {\rm Gene Tsudik}\\
    University of California Irvine\\
    gts@ics.uci.edu
}
	
\IEEEoverridecommandlockouts
\makeatletter\def\@IEEEpubidpullup{6.5\baselineskip}\makeatother
\IEEEpubid{\parbox{\columnwidth}{
		Symposium on Usable Security and Privacy (USEC) 2026 \\
		27 February 2026, San Diego, CA, USA \\
		ISBN 978-1-970672-07-7 \\
		https://dx.doi.org/10.14722/usec.2026.23017 \\
		www.ndss-symposium.org, https://www.usablesecurity.net/USEC/
}
\hspace{\columnsep}\makebox[\columnwidth]{}}

\maketitle

\begin{abstract}
Multi-Factor Authentication (MFA) enhances login security by requiring users to use multiple 
authentication factors. MFA adoption has surged in recent years in response to the growing
frequency, diversity, and sophistication of attacks. 
{\em Duo} is among the most popular MFA providers, used by thousands of organizations worldwide, 
including Fortune 500 companies and large educational institutions. However, its usability
has not been investigated thoroughly or recently. Although prior work addressed 
technical challenges and user perceptions during initial implementation phases, there was
no assessment of key usability metrics, such as average task completion time 
and System Usability Scale (SUS) scores. Moreover, relevant prior results are outdated, 
having been conducted years ago when the entire MFA concept was relatively new and unfamiliar 
to the average user.

Motivated by the above, we conducted a long-term and large-scale Duo usability 
study. This study took place at the University of California Irvine (UCI) over the course 
of the 2024-2025 academic year and it involved $2,559$ unique participants. Our analysis is 
based on a large set of authentication log files and a survey of $57$ randomly selected participants. 
The study reveals that the average overhead of a Duo Push notification task is nearly
$8$ seconds, a duration described by participants as short to moderate. Several factors 
influence this overhead, including the time of day when the task was performed, the 
participant's field of study, as well as their education/student level. The rate of authentication 
failures due to incomplete Duo tasks is $4.35$\%. Furthermore, $43.86$\% of survey respondents 
reported experiencing a Duo login failure at least once. The Duo SUS score is found to be $70$,
corresponding to a ``Good'' usability level: while participants generally find Duo easy to use, 
they also perceive it as annoying. On a positive note, Duo increases participants' sense 
of security regarding their accounts. Finally, participants described commonly encountered
issues and provided constructive suggestions for improvement.

\end{abstract}

\IEEEpeerreviewmaketitle
\section{Introduction}
The increasing sophistication and diversity of attacks, coupled with the limitations of traditional 
single-factor (password-based) authentication, motivate the widespread adoption of so-called 
Multi-Factor Authentication (MFA) techniques \cite{aloul2009two}. Though the trend towards MFA adoption
is not new, it has accelerated in the last $5$-$10$ years. While conventional 
password-based authentication remains vulnerable to phishing, key-logging, and brute-force attacks
\cite{das2014tangled}, MFA appreciably reduces the efficacy of such attacks. 
As of 2025, $\approx~98\%$ of organizations worldwide support MFA, with large enterprises having
an MFA adoption rate of $87$\% \cite{mfa_adoption}. Also, nearly two-thirds of individual users now 
use MFA \cite{mfa_users}, which confirms its growing importance.

The concept of MFA is quite broad. The only unifying factor is that it must involve two or more
modalities or factors, including: what you know (passwords or PINs), what you recognize (answers to 
personal questions with multiple-choice answers or pictures), what you are (biometrics), and what you have 
(various types of tokens). Although there are many combinations of these factors, most 
civilian/consumer settings use Two-Factor Authentication (2FA), with the primary (first) factor
based on traditional passwords. The second factor is usually either biometrics or tokens.

Among 2FA providers, Duo is one of the most popular, recognized for its reliability, ease of use, 
and strong security features. Duo ranks among the top MFA providers globally, and is considered by some metrics
to be the best 2FA app \cite{duo_rank}. Although Duo does not publicly disclose the total number of its users,
it is adopted by many major corporations, including Fortune 500 companies, e.g., Microsoft, Walmart, IBM, and 
Cisco \cite{fortune_500_duo_users}. Similarly, many large educational entities (e.g., 
University of Illinois \cite{uiuc_duo}, University of Florida \cite{uf_duo}, University of California, Berkeley 
\cite{ucb_duo}, University of Arizona \cite{uarizona_duo}, and Columbia University \cite{columbia_duo}) have integrated 
Duo 2FA into their user authentication frameworks.

As technology evolves and users gradually become more accustomed to MFA, it is important to assess whether 
Duo meets current user needs and expectations. Despite its popularity, the usability of Duo remains underexplored. 
To the best of our knowledge, no previous research has assessed the overhead of Duo in terms of task completion time 
or determined Duo's System Usability Scale (SUS) score.  Prior studies, e.g., ~\cite{reynolds2020empirical} and 
\cite{weidman2017like}, focused on technical challenges and user perception during the initial rollout of Duo, 
when its use was newly mandated within organizations. In these initial and transitional contexts, user attitudes 
were potentially influenced by disruptions to established authentication routines. For example, users accustomed 
to single-factor authentication might have perceived Duo as an added burden, whereas new users introduced to Duo 
from the outset likely viewed it as a normal part of the authentication process. Other prior research on the 
overhead and usability of push notification methods in 2FA systems~\cite{acemyan20182fa,reese2019usability}, 
conducted in 2017 and 2018, respectively, is simply outdated. Back then, 2FA technology was relatively new 
and users had not yet integrated 2FA processes into their authentication routines. 

Given these limitations, we aim to gain a fresh perspective by examining Duo in a context where users are 
familiar with 2FA and encounter it on a regular basis. Moreover, by focusing on routine use rather than 
initial adoption, we can more accurately evaluate Duo’s usability without any biasing effects introduced during 
transitional phases. To this end, we analyzed authentication logs 
collected over nine months, involving $2,559$ unique users, most of whom were students at the University of California, 
Irvine (UCI). These logs contain timing events that allow us to 
compute the overhead of Duo push notification tasks and the authentication failure rate. 
We also conducted a post-study survey -- with randomly chosen
$57$ participants -- to gain insights into user experiences and perceptions.

\noindent
Our work addresses the following research questions:

\noindent \textbf{RQ1. What is the overhead of the Duo push notification task?}
This overhead is determined by the task completion time. 
Completion means a user clicking the ``Approve'' or ``Deny'' button in a push notification or in the Duo app.
On average, users take $7.82$ seconds, which is generally perceived as short-to-moderate time. 

\noindent \textbf{RQ2. What factors influence the overhead of a Duo push notification task?}
Perhaps unsurprisingly, this turns out to be influenced by the time of day. 
Additional factors include the participant's major/field of study and their student/education level.

\noindent \textbf{RQ3. How likely are users to experience authentication failures when using Duo?}
Incomplete Duo tasks lead to a failure rate of $4.35\%$. Also, over 
$43.86\%$ of survey respondents reported experiencing at least one failed Duo login.

\noindent \textbf{RQ4. What do users think about Duo?}
Most users express positive or mixed perceptions. Many find it annoying, however, they 
acknowledge its security benefits. Users note issues, such as problems with push notifications and complications 
when changing phone numbers or devices. They also provide constructive suggestions for improving the overall
user experience.

\noindent \textbf{RQ5. What is the SUS-based usability of Duo?}
Duo earns a ``Good'' usability level, with an average SUS score of $70$. While Duo tasks are considered 
``easy'', most users also view them as ``annoying''.

Through the combination of authentication log analysis and post-study survey, we comprehensively assess Duo
usability and user perceptions. We believe that this provides a timely and expanded understanding of the Duo ecosystem. 

\noindent {\textbf{Outline:}} Section~\ref{sec:background} provides some background on MFA, and SUS. 
Section~\ref{sec:user_study} describes the methodology of the user study. Section~\ref{sec:eval} presents 
the results and their analysis. Section~\ref{sec:related_work} compares our results to previous user 
studies. Section~\ref{sec:discussion} discusses the limitations of this study and provides recommendations 
to improve Duo MFA usability. Finally, Section~\ref{sec:conclusion} concludes the paper.

\section{Background} \label{sec:background}
\subsection{MFA}
MFA is a security mechanism that mandates two or more verification factors for user authentication:
\begin{compactitem}
    \item \textbf{Knowledge}: Something the user knows, such as a password, PIN, or security question.
    \item \textbf{Cognition}: Something the user recognizes, such as familiar faces or answers to personal questions.
    \item \textbf{Possession}: Something the user has, such as a smartphone, security token, or smart card.
    \item \textbf{Inherence}: Something the user is (aka biometrics) based on, e.g., fingerprints.
\end{compactitem}
The most common form of MFA is 2FA, which requires two factors. 

One of the earliest forms of 2FA is the use of Automated Teller Machines (ATMs), which combines the possession of an ATM
card and the Knowledge of a secret PIN \cite{jain2004introduction}. More recent 2FA instantiations involve combining a
username/password with either (1) one-time PINs or codes sent via SMS, email, or generated by specialized authentication apps,
or (2) biometric verification, such as fingerprint or facial recognition.

Growing adoption of 2FA is driven by regulatory requirements and industry standards that mandate strong(er) 
authentication protocols. For example, the Payment Card Industry Data Security Standard (PCI DSS) and the General 
Data Protection Regulation (GDPR) emphasize the importance of robust authentication measures.

Despite its advantages, 2FA is not without challenges. Usability concerns, such as the complexity of
managing multiple authentication factors and potential accessibility issues for users with disabilities, 
must be addressed to ensure widespread adoption \cite{bonneau2012quest}. Also, the security of MFA systems 
can be compromised if not properly implemented, since attackers naturally target the weakest link in the 
authentication chain.

\subsection{SUS}
\label{sec: sus}
The System Usability Scale (SUS) \cite{brooke1996sus} is a time-tried, widely recognized, and 
efficient survey method for evaluating the usability of various products and systems. It consists 
of a standard set of ten questions (shown in Table \ref{tab:sus-questions}) where 
odd-numbered questions are positively phrased, and even-numbered questions are negatively phrased. 

\begin{table}[h]
\caption{Standard SUS Questionnaire}
\scriptsize
\label{tab:sus-questions}
\centering
 \begin{tabularx}{\linewidth}{p{0.1cm} p{12.5cm}}
\toprule  
\textbf{No} & \textbf{Question} \\ \midrule
 1 & I think that I would like to use this system frequently. \\
 2 & I found the system unnecessarily complex. \\
 3 & I thought the system was easy to use. \\
 4 & I think that I would need the support of a technical person to use this system. \\
 5 & I found the various functions in this system were well integrated. \\
 6 & I thought there was too much inconsistency in this system. \\
 7 & I would imagine that most people would learn to use this system very quickly. \\
 8 & I found the system very cumbersome to use. \\
 9 & I felt very confident using the system. \\
10 & I needed to learn a lot of things before I could get going with this system. \\
\bottomrule
\end{tabularx}
\end{table}

Each question offers five response options: Strongly Disagree, Disagree, Neutral, Agree, and Strongly Agree. 
A numerical SUS score is derived from an individual's responses to these questions, with the score ranging 
from 0 to 100. The procedure for computing a SUS score includes the following steps:
\begin{compactitem}
    \item Assign numerical values to responses: Strongly Disagree = $1$, Disagree = $2$, Neutral = $3$, Agree = $4$, and Strongly Agree = $5$
    \item For odd-numbered questions, subtract 1 from the corresponding response value 
    \item For even-numbered questions, subtract the response value from $5$
    \item Sum up all resulting scores and multiply the total by $2.5$ to obtain the final SUS score
\end{compactitem}

The average SUS score is computed by averaging individual SUS scores of all participants. 
Adjective scaling \cite{bangor2009determining} is then applied to interpret the absolute
usability level of the product from the average SUS score. This scaling includes seven 
usability levels, ranging from ``worst imaginable'' to ``best imaginable'' usability. 
Table \ref{tab: sus adjective scaling} shows the relationship between SUS score and usability levels.

\begin{table}
\caption{Adjective Ratings of SUS Scores}
\scriptsize
\label{tab: sus adjective scaling}
\centering
 \begin{tabularx}{\linewidth}{p{2.5cm} p{2.5cm} p{2.5cm}}
\toprule  
  \textbf{Adjective} & \textbf{Mean SUS Score} & \textbf{SUS Score Range} \\
  \midrule
  Worst Imaginable & 12.5 & 0 - 25 \\
  Poor & 20.3 & 25.1 - 39 \\ 
  OK & 35.7 & 39.1 – 51.6 \\ 
  Good & 71.4 & 51.7 - 73.5 \\ 
  Excellent & 85.5 & 73.6 - 85\\ 
  Best Imaginable & 90.9 & 85.1 - 100 \\
\bottomrule
\end{tabularx}
\end{table}

\section{Methodology} \label{sec:user_study}

\subsection{Setting}
The study was conducted over nine months: from August 2024 to April 2025. It took place at UCI, 
in the Donald Bren School of Information and Computer Sciences.
The school's IT department manages a Linux server cluster used by faculty, students, and staff. 
Access to this cluster is required for many courses. Students from all schools on campus
who enroll in these courses have access to it. Duo serves as the second factor
for user authentication to access the Linux cluster. User authentication proceeds as follows:
\begin{figure}[h]
    \centering
    \includegraphics[width=\columnwidth]
    {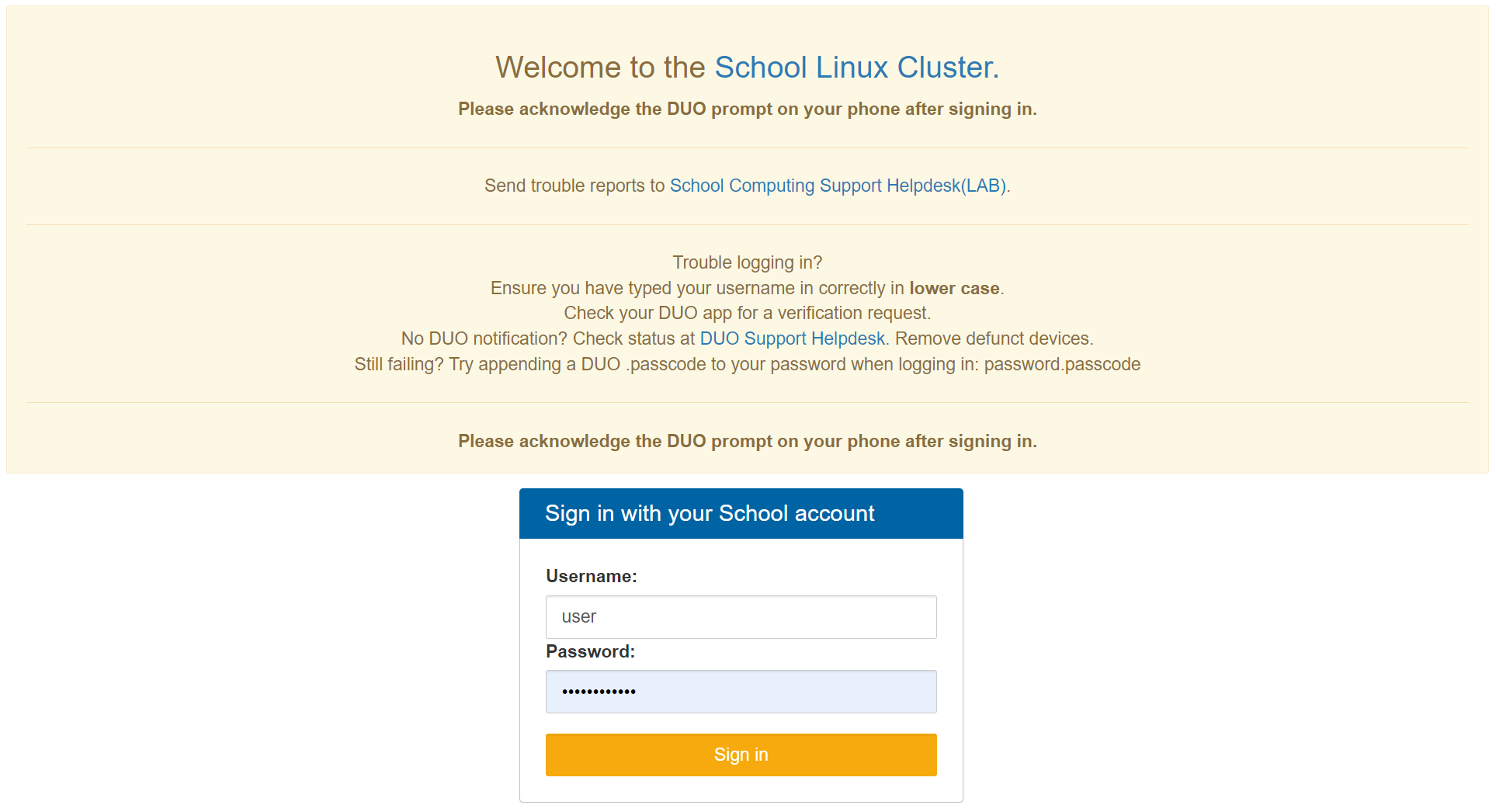}
    \caption{Sign-In Page for Linux Cluster}
   
    \label{fig:signin_page}
\end{figure}

\begin{figure}[h]
    \centering
    \includegraphics[width=\columnwidth]
    {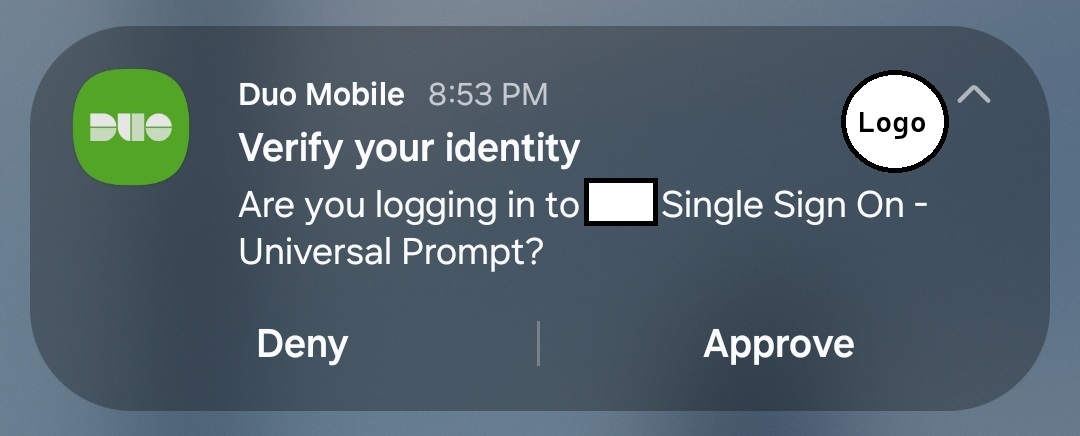}
    \caption{Duo Push Notification}
   
    \label{fig:duo push}
\end{figure}

\begin{compactenum}
    \item \textbf{First Factor:} Entering one's userid and password (sign-in in Figure \ref{fig:signin_page}).
    Successful verification of the userid and password prompts Duo to send
    a push notification to the mobile device registered for the supplied userid.
    \item \textbf{Second Factor:} Pressing the {\tt Approve} or {\tt Deny} button in the Duo app push notification. 
    Figure \ref{fig:duo push} shows an example.
\end{compactenum}
In some cases, a user cannot complete the Duo task through the push notification, e.g., 
if they clear the notification or the button on the push notification does not work. 
In such cases, the user opens the Duo app and presses the desired button. 

\begin{figure}[h]
    \centering
    \includegraphics[width=0.6\columnwidth]
    {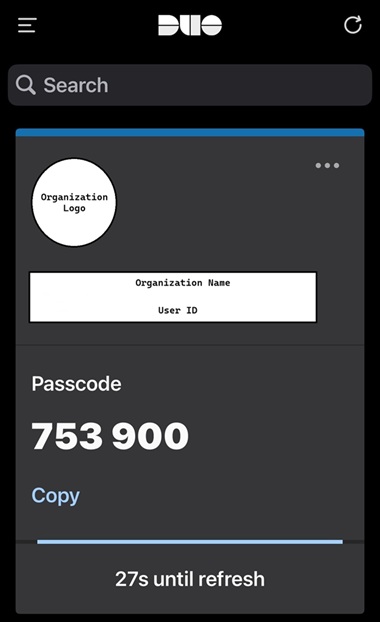}
    \caption{Passcode generated inside Duo App}
   
    \label{fig:duo code}
\end{figure}

Users without a smartphone use physical tokens. They can obtain a one-time passcode from the 
physical token and append it to the password in the ``password.passcode" format.
In certain situations, users with smartphones must also provide a Duo one-time passcode,
instead of, or in addition to, pressing the approve button on the Duo push notification. 
Figure \ref{fig:duo code} shows an example of a one-time passcode generated in the Duo app. 

Duo log files were duly obtained from the school's IT staff. Each sign-in instance has either one 
or two log entries. For each failed primary factor (userid/password-based) authentication attempt, 
there is a single timestamped log entry. If the primary factor authentication succeeds, there are two 
timestamped log entries: (1) for the primary factor, and (2) for either a successful or failed second factor. 
Each log entry contains a userid, which, along with the timestamp, links (1) and (2). 

Although log files contained entries for two types of second factor authentication (Duo push 
notification and passcode), we can only reliably measure the overhead of push notification. 
It reflects how long a user takes to click on the Approve or Deny button. Because a user 
appends the passcode to the password, measuring the time spent retrieving the passcode from the 
user's device is infeasible. Log entries for primary and secondary  
authentication occur within milliseconds of each other and reflect the server's passcode 
verification time.

\subsubsection{Website}
Faculty, staff, and students use an internal website to access the online Linux cluster. 
Access to that website is restricted to individuals using the campus network or a 
Virtual Private Network (VPN). Figure \ref{fig:signin_page} shows a snapshot of the sign-in
page for this website,  which has been slightly modified to preserve the anonymity of the authors.

\subsubsection{Directory Crawler}
UCI has a public directory that includes information on students, faculty, and staff. This 
directory contains non-sensitive details, such as names, userids, and email addresses for all 
university personnel. For students, the directory also provides major and student level information. 
For staff and faculty, it includes titles. To gather relevant information about study participants, 
we used a crawler specifically designed to collect all this information. This crawler uses a userid
extracted from a log file as a search key.

\subsection{Logistics and Data Cleaning}
We analyzed a total of $96,048$ log entries from which we extracted $2,804$ unique userids. 
Some of these extracted userids were invalid as a result of user typos. (Naturally, they all
corresponded to failed Duo authentication attempts.) To validate userids, we used the university's 
public directory -- a userid was considered valid  if it was found in the directory. This process
yielded $2,561$ valid userids. $552$ log entries corresponding to $243$ invalid userids were 
removed. 

Furthermore, over half of the log entries ($49,942$) lacked a userid and were thus excluded 
from the Duo overhead calculation, since we could not create a primary authentication log entry, 
Duo authentication log entry pair from them. However, we included these entries in our analysis 
to determine the rate of authentication failures caused by Duo. This is because some 
log entries with a ``Primary authentication successful'' message did not have any userid. 
However, they should still be included in the number of successful primary authentications.  

Among the $2,561$ valid userids, three were removed. One belonged to 
one of the authors and the other two were from participants who opted out. 
$13$ log entries corresponding to these three userids were also excluded from the analysis. 
In the end, the analysis of Duo overhead focused on $45,541$ log entries corresponding to $2,558$ userids.

\subsection{Survey}
We invited $800$ randomly selected participants to complete a post-study survey, of whom $57$
responded and completed the survey.  
On average, participants took about $5$ minutes to finish the survey, and each was compensated with 
a \$$5$ Amazon gift card. The survey's purpose was to gather comprehensive feedback on several 
aspects of Duo use. It included questions about participants' overall perceptions of Duo, 
specifically their opinions about the time needed to complete Duo-related tasks and 
whether they were perceived as easy, difficult, fun, or annoying. Furthermore, the survey 
sought to understand whether participants believed that Duo improves the security of their accounts. 
Participants were also asked about their experiences of Duo authentication failures. To assess 
Duo usability, the survey incorporated SUS questions. 
The complete Survey questionnaire is included in the Appendix. 

\subsection{Ethical Considerations}
This study received approval from UCI Institutional Review Board (IRB).
No additional steps were added to the normal user workflow due to our study, i.e., 
no extra burden was imposed on the participants. All information collected using the 
directory crawler is publicly available and 
does not compromise participants' privacy. The IRB granted permission to use the authentication log 
files for research purposes without requiring explicit consent from participants. Nevertheless, 
to err on the side of safety and ethics, we informed all participants about the intended use of their 
authentication log data and provided them with the opportunity to express any concerns. Out of 
$2,561$ notified participants, only $4$ responded: $2$ expressed curiosity about the 
study and were provided with additional information, while another $2$ requested to opt out, 
and we removed their data from the analysis. For the second phase of the study, which included the 
survey, participants provided explicit consent by completing an online consent form. 
Access to the survey questionnaire was granted only after consent was obtained.
\section{Results and Analysis} \label{sec:eval}
\subsection{Demographics}
Not surprisingly, almost all participants were students. The exact breakdown was: 
$2,521$ students, $9$ faculty members, and $28$ staff. Most students were undergraduates: 
$1,265$ seniors, $801$ juniors, $248$ sophomores, and $7$ freshmen. There were also 
$200$ graduate students.

Students represented 63 academic majors. However, the highest numbers corresponded to five 
majors related to Computer Science: $1,417$ in Computer Science (CmptSci), $251$ in Software Engineering 
(SW Engr), $208$ in Computer Science and Engineering (CSE), $125$ in Data Science (DataSci), and 
$111$ in Informatics (IN4MATX).

Although we lack specific information regarding participants' age, gender, and highest 
level of degree attained (unavailable in the public directory), reasonable assumptions 
can be made based on their academic status. Given that most participants are undergraduates, 
their ages very likely range between $18$ and $22$. All participants have at least a high school 
diploma. Faculty participants are likely to hold PhD degrees, while staff and graduate students 
typically have at least a bachelor's degree. Age distribution is estimated to be approximately 
$90.73\%$ aged $18-22$ and $9.27\%$ over $22$. In terms of highest degree attained, 
the breakdown is estimated at: $90.74\%$ -- high school, $8.91\%$ -- bachelors, 
and $0.35\%$ -- PhD.

\begin{figure}[h]
    \centering
    \includegraphics[ width=\columnwidth]
    {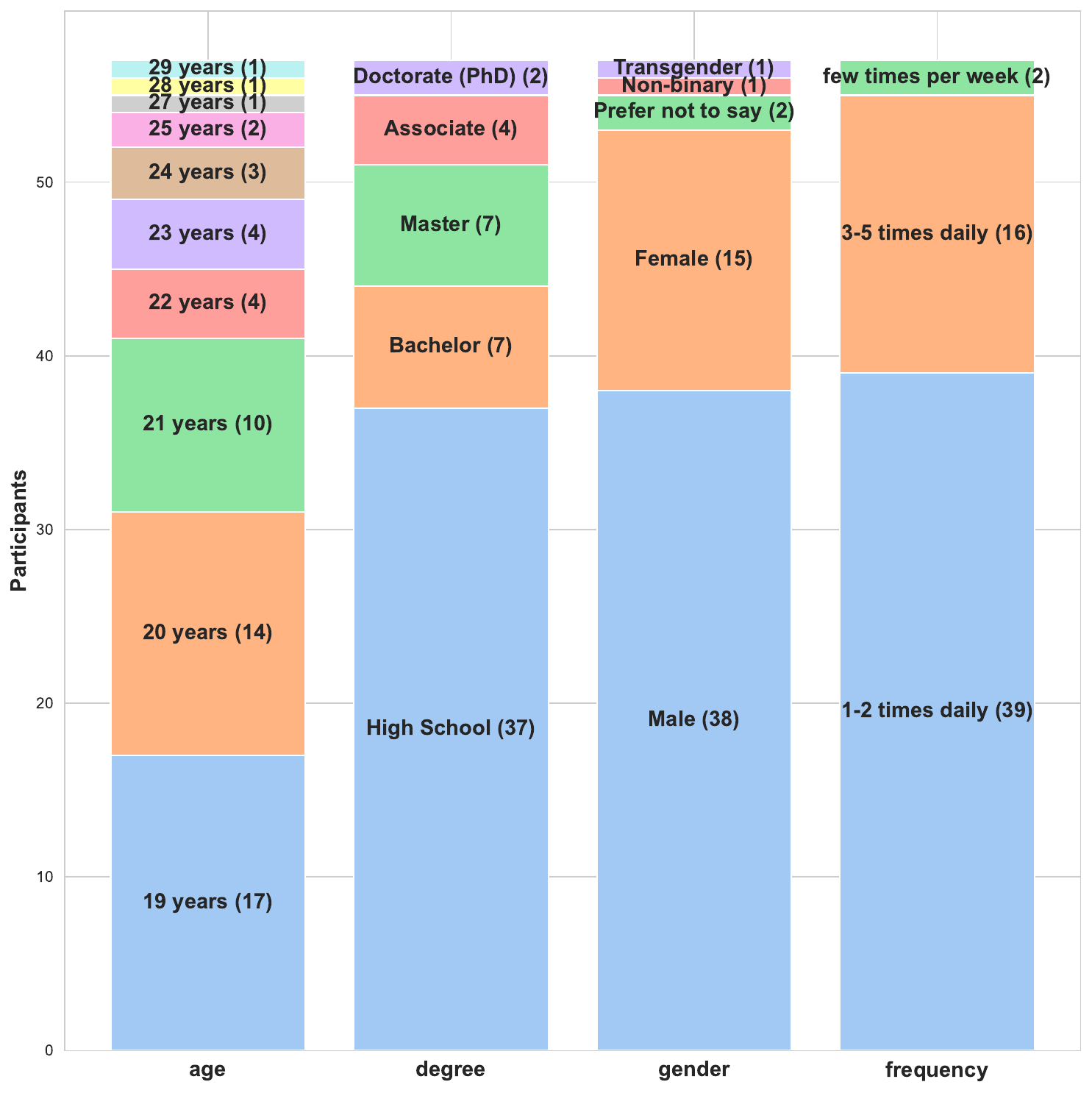}
    \caption{Demography of Post Survey Participants}
    
    \label{fig:survey demography}
\end{figure}

Fortunately, we could obtain detailed demographic information from participants who completed 
the survey. Figure \ref{fig:survey demography} shows the demographics. Gender-wise, 
two-thirds were male. Also, two-thirds had a high school diploma as the highest 
educational degree attained, indicating current undergraduate status.
In terms of age, all participants were $19$ to $29$, with the majority
($41$) between $19$ and $21$. Participants were also asked about their frequency of
using Duo: most use it $1-2$ times per day.

\subsection{Qualitative Analysis of Open‑Ended Responses}
We analyzed all open‑ended survey responses using thematic analysis. Two researchers independently 
performed inductive coding on the responses, iteratively grouping codes into higher‑level themes. 
Disagreements were resolved through discussion. The themes reported in subsequent sections reflect 
the most salient issues raised by participants.

\subsection{Duo Push Overhead}
We define Duo push overhead as the elapsed time between a log entry for successful 
primary authentication and the subsequent log entry for successful Duo authentication. 
Each log entry provides detailed information, i.e., fields, including: userid, timestamp, 
and message. The userid helps us isolate log entries 
associated with individual users, subsequently sorting these entries in chronological order 
based on timestamps. The message field allows us to identify pairs of log entries where a successful 
primary authentication entry is immediately followed by a successful Duo authentication entry.
The elapsed time between two such entries is the Duo overhead for a specific sign-in instance. 

This process yielded data for $16,851$ instances.
The average Duo push overhead across all instances is $9.29$ seconds. The median is $6.89$ seconds, with the first quartile (Q1) at $4.82$ seconds, the third quartile (Q3) at $10.47$ seconds, and an interquartile range (IQR) of $5.65$ seconds. However, the distribution exhibits a pronounced long tail, as described in Section \ref{sec: overhead distribution}. To mitigate the effect of outliers, we discarded values above the 95th percentile (i.e., over $24.71$ seconds). After trimming, the average overhead decreases to $7.82$ seconds, the median to $6.67$ seconds, Q1 to $4.73$ seconds, Q3 to $9.61$ seconds, and the IQR to 4.00 seconds. We hereafter refer to the revised average as the average Duo push overhead.

The comparison between values before and after trimming demonstrates that trimming substantially reduces the mean and narrows the spread, while leaving the median and quartiles largely intact. Thus, trimming improves the validity of the reported average by reducing the impact of outlier values, without altering the central tendency of the distribution.

\subsubsection{Distribution} \label{sec: overhead distribution}
\begin{figure}[h]
    \centering
    \includegraphics[width=\columnwidth]
    {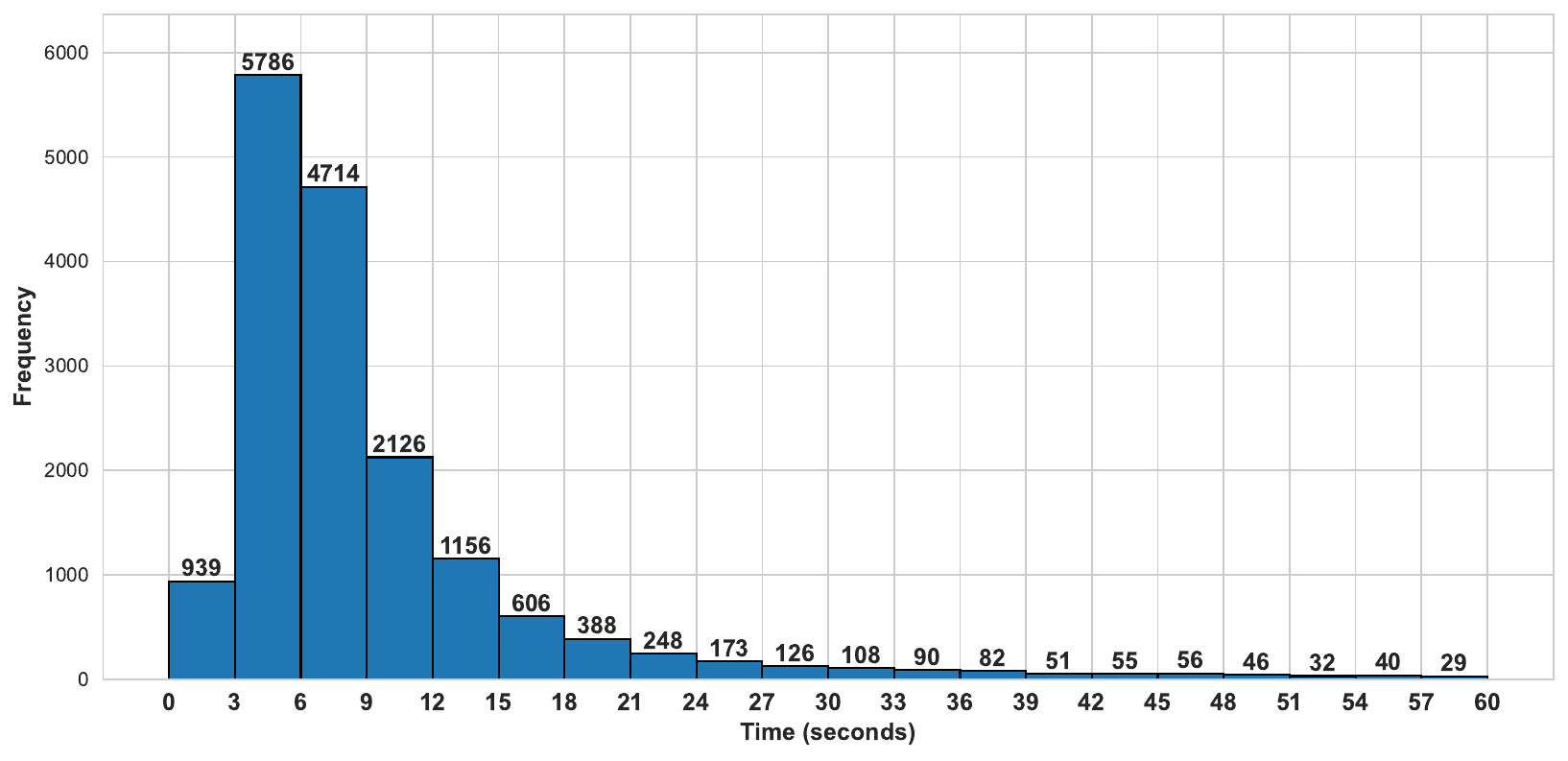}
    \caption{Histogram of Duo Push Overhead}
    \label{fig:push_overhead_histogram}
\end{figure}

Figure \ref{fig:push_overhead_histogram} 
illustrates the distribution of push overhead across all instances. The minimum and maximum  
are $1.15$ and 59.4 seconds, respectively. Most overhead values fall between $3$ and $15$ 
seconds. Recall that this overhead corresponds only to successful sign-ins and is limited by 
the Duo push notification expiration time, set to $1$ minute in our experimental setting.

\subsubsection{Statistical Tests}
Since multiple Duo authentication attempts can originate from the same participant, the 
resulting push overheads are not independent across all observations. They are instead 
clustered within individuals. This motivates the use of a linear mixed‑effects model that 
explicitly accounts for repeated measurements from the same participant. We partition push 
overheads into distinct groups based on participants’ majors, student levels, and times of day 
when the Duo tasks were performed. To ensure that model estimates are based on sufficiently 
representative data, we restrict each analysis to groups that include at least $20$ unique 
userids; groups that do not meet this criterion are excluded from the corresponding comparisons.

For each grouping factor (major, student level, time of day), we fit a linear mixed‑effects model with Duo push overhead as the dependent variable, the grouping factor as a fixed effect, and a random intercept for each participant to capture within‑subject correlation. We then use this model to test whether there are statistically significant differences among the levels of the factor. When an overall effect is detected, we perform post‑hoc pairwise comparisons between group levels using contrasts of the model‑estimated marginal means. To control for inflated Type I error due to multiple pairwise tests, we apply the Benjamini–Hochberg correction to the resulting p‑values before determining which pairs are statistically significant. Only those pairs with corrected p‑values below the chosen significance threshold of $0.05$ are reported as exhibiting significant differences. 

\subsubsection{Time of Day}
\begin{figure}[h]
    \centering
    \includegraphics[width=\columnwidth]
    {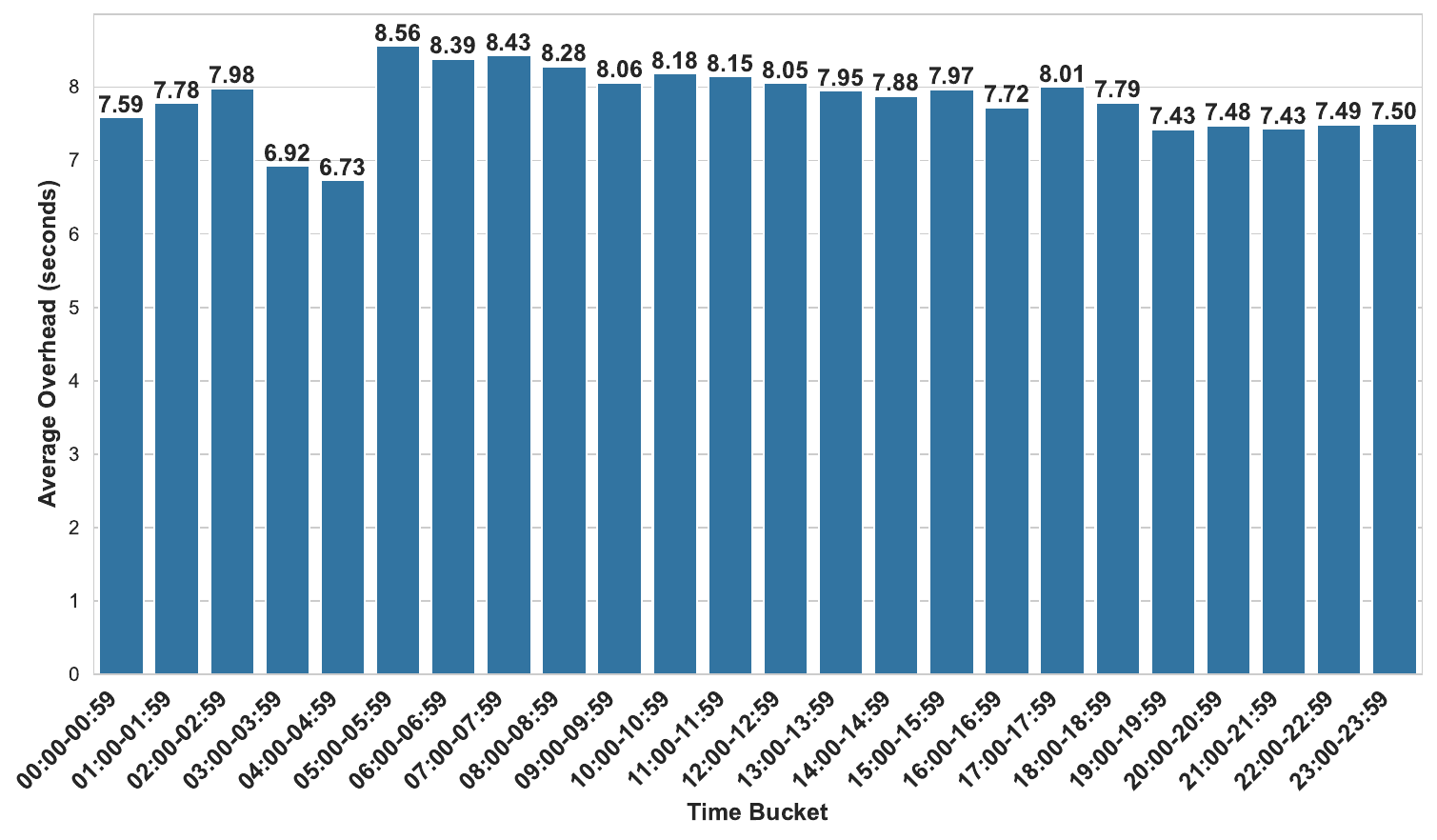}
    \caption{Average Duo Push Overheads for Different Times of the Day}
    
    \label{fig:push_overhead_bucket}
\end{figure}

We partitioned sign-in instances into $24$ groups, corresponding to each hour of the day. 
We then computed the average Duo push overhead for each group, as shown in Figure 
\ref{fig:push_overhead_bucket}. Timestamps in log files were initially recorded as 
UTC and we converted them into the local time zone for this analysis. Two notable patterns 
emerged: (1) an overall decreasing trend throughout the day: from $5$ a.m. to $5$ p.m., and (2) an overall increasing 
trend from midnight to $3$ a.m. Curiously, the two lowest average timings occurred late at night ($3$ a.m.--$5$ a.m.), 
followed immediately by the four highest averages ($5$ a.m.--$9$ a.m.).

The analysis revealed statistically significant differences in overheads based on the time of day. We identified 43 different hourly pairings with significant differences.

\subsubsection{Student Level}
We examined the effect of student level on overhead (Table \ref{tab: student level overhead}). Average overheads increase from freshmen ($6.88$ seconds) to graduate students ($8.64$ seconds). Undergraduate medians are similar ($6.47$–$6.81$ seconds), while graduate students have a higher median ($7.46$ seconds). Quartiles show a similar pattern: graduate students have higher Q1 ($5.53$ seconds) and Q3 ($10.41$ seconds), compared to undergraduates (Q1 $\approx$ $4.5$–$4.7$ seconds, Q3 $\approx$ $9.5$ seconds). Variability also differs: freshmen have the smallest IQR ($3.74$ seconds), sophomores the largest ($5.11$ seconds), and graduates remain relatively wide ($4.89$ seconds). We observed statistically significant differences in overheads between graduate and (sophomore,
junior, and senior) undergraduate students.

\begin{table}[h]
\caption{Duo Push Overheads per Student Level (seconds)}
\label{tab: student level overhead}
\centering
 \begin{tabularx}{\linewidth}{cccccc}
\toprule  
  Student Level & Mean & Median & Q1 & Q3 & IQR \\
  \midrule
  Freshmen & 6.88 & 6.81 & 4.49 & 8.23 & 3.74 \\
  
  Sophomore & 7.71 & 6.47 & 4.55 & 9.66 & 5.11 \\ 
  
  Junior & 7.74 & 6.60 & 4.65 & 9.51 & 4.86 \\ 

  Senior & 7.76 & 6.62 & 4.72 & 9.47 & 4.75 \\ 
  
  Graduate & 8.64 & 7.46 & 5.53 & 10.41 & 4.89 \\ 

\bottomrule
\end{tabularx}
 
\end{table}

\subsubsection{Effect of Major}
\begin{figure}[h]
    \centering
    \includegraphics[width=\columnwidth]
    {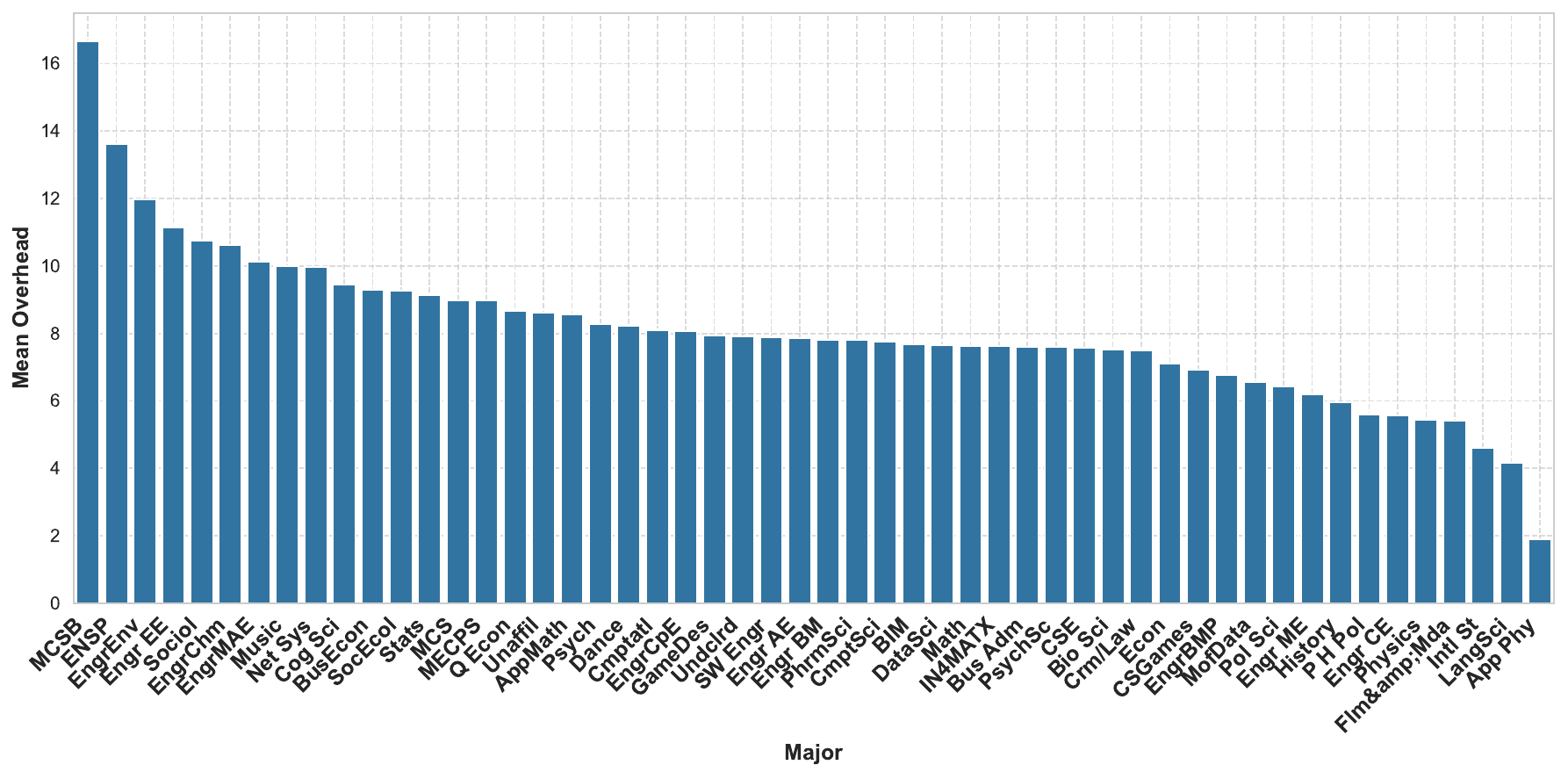}
    \caption{Average Duo Push Overheads for Different Majors}
    
    \label{fig:push_overhead_major}
\end{figure}

Duo push overhead is influenced by participants' academic major (field of study), 
as shown in Figure 
\ref{fig:push_overhead_major}. We identified $8$ pairs of majors that exhibit statistically 
significant differences in push overheads:

\begin{enumerate}
    \item Master's in Computer Science (MCS), Computer Science (CmptSci)
    \item Master's in Computer Science (MCS), Computer Science and Engineering (CSE)
    \item Master's in Computer Science (MCS), Data Science (DataSci)
    \item Master's in Computer Science (MCS), Software Engineering (SW Engr)
    \item Undeclared (Undclrd), Data Science (DataSci)
    \item Undeclared (Undclrd), Computer Science (CmptSci)
    \item Undeclared (Undclrd), Computer Science and Engineering (CSE)
    \item Master's in Computer Science (MCS), Applied Math (AppMath)
\end{enumerate}

In particular, MCS appears in the top four pairs. This major consists exclusively of professional master’s students. In contrast, CSE, DataSci, BIM, Undclrd, and AppMath consist entirely of undergraduate students. Among participants with the SW Engr major, $96.41$\% are undergraduates, closely followed by CmptSci, with $94.7$\% undergraduates.

These findings further support the earlier observation that overhead is influenced by the
student level.  

\subsection{Perceptions of Using Duo}
We surveyed participants to gather their perceptions of Duo, using both open-ended and 
multiple-choice questions. The former were posed first to minimize potential bias 
introduced by multiple-choice options.

Participants were asked separately about Duo push notification tasks and code input tasks, 
since the latter is generally more complex and time-consuming than the former. Considering that 
only some participants experienced the code input task, everyone was initially asked whether they 
had ever encountered it during the sign-in process. Only those who answered affirmatively were 
subsequently presented with questions about the code input task.

\subsubsection{Multiple-Choice Responses}
\begin{figure}[h]
    \centering
    \begin{subfigure}{\columnwidth}
        \centering
        \includegraphics[width=\columnwidth]{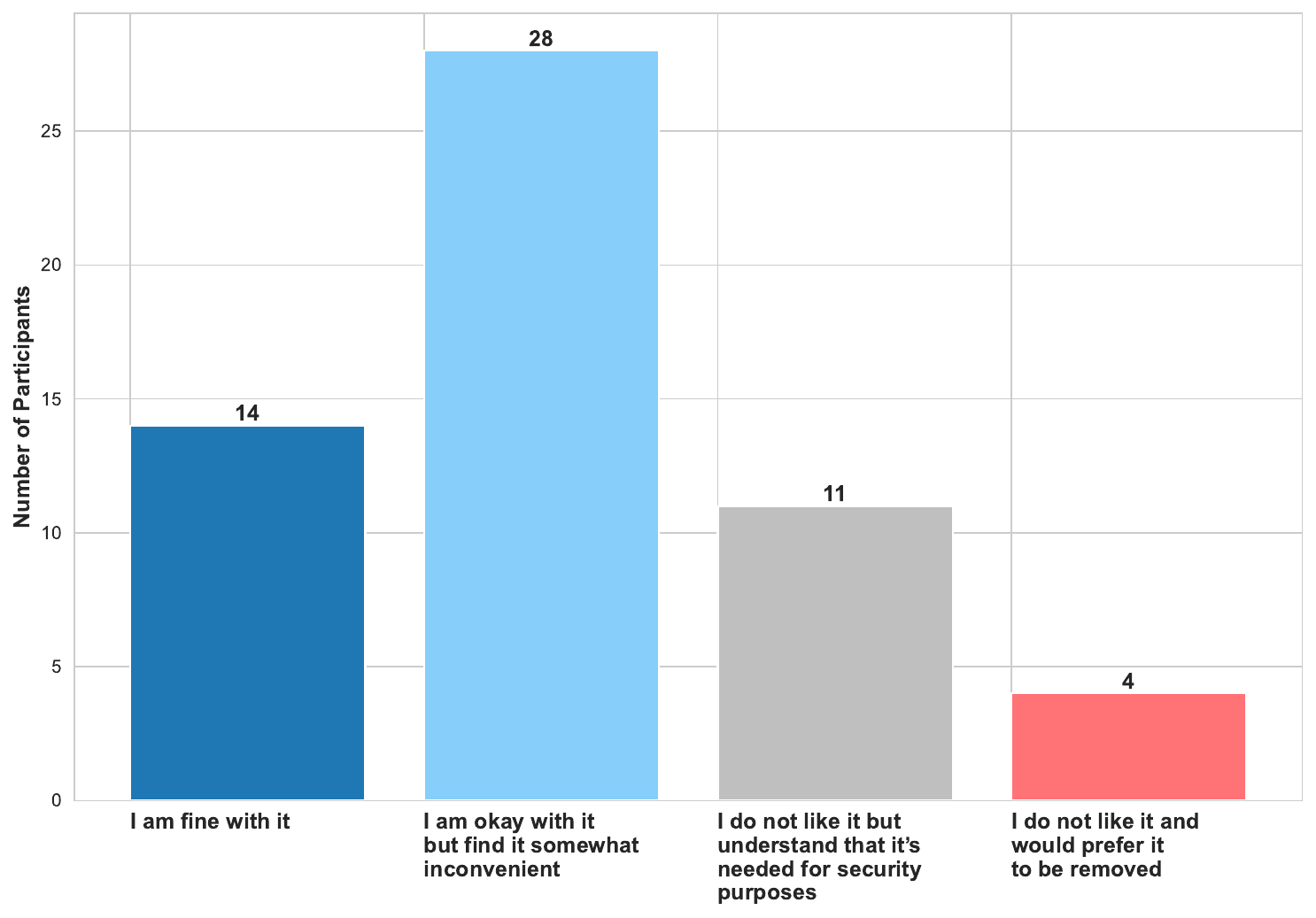}
        \caption{Duo Push Notification Task}
    \end{subfigure}
    \begin{subfigure}{\columnwidth}
        \centering
        \includegraphics[width=\columnwidth]{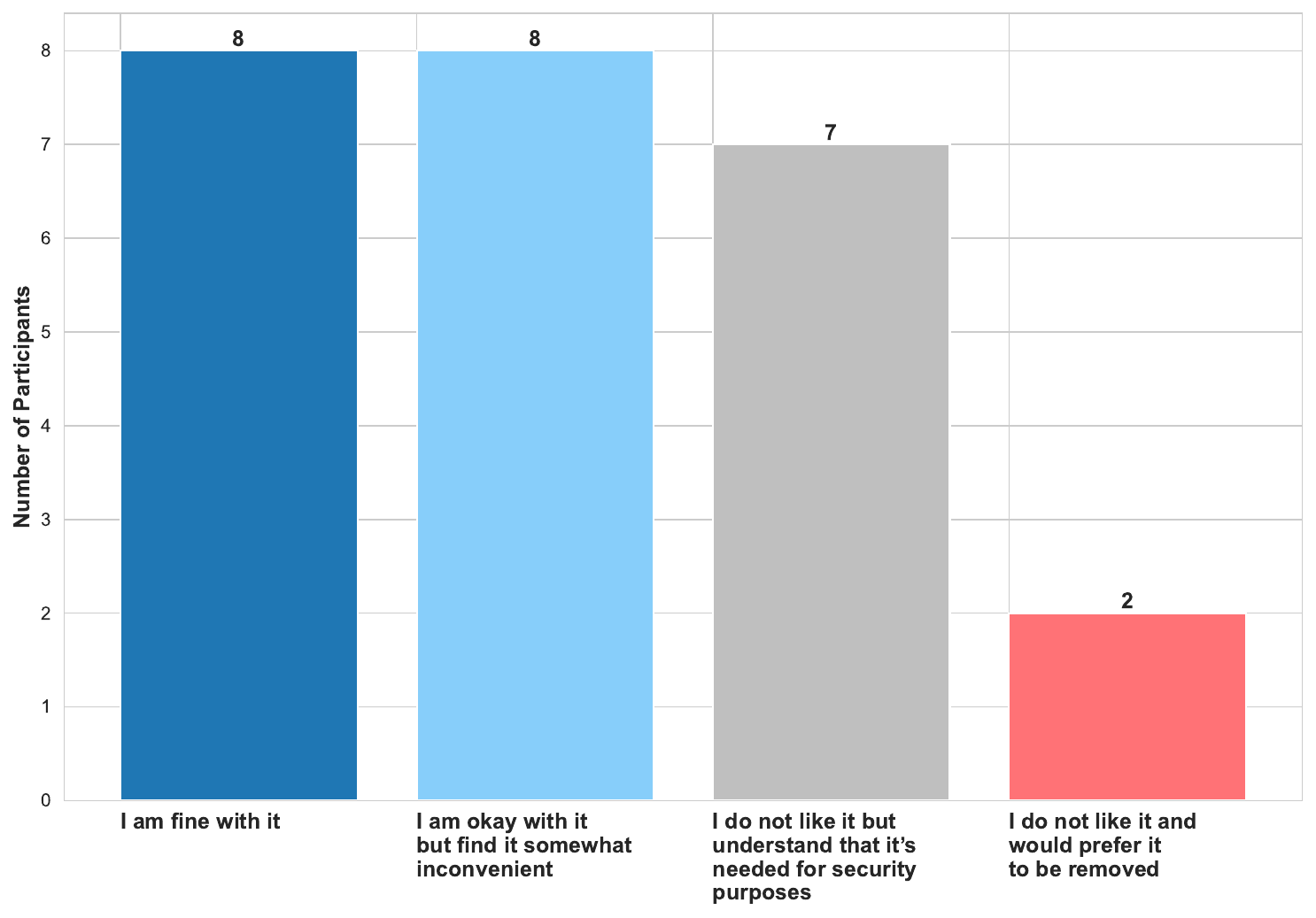}
        \caption{Duo Code Input Task}
    \end{subfigure}
    \caption{Multiple-Choice Responses: Perceptions of Duo}
    \label{fig:push_code_opinion}
\end{figure}

Multiple-choice questions for both the push notification task and the code input task 
included identical options.   
Figure \ref{fig:push_code_opinion} presents the number of participants who selected each option. 
For the push notification task, $73.68$\% of participants indicated they were either fine or 
okay with minor inconveniences. Only $7.01$\% expressed a desire to have it removed, 
whereas the remaining $19.31$\% acknowledged that it is necessary for security reasons, despite not 
liking it. For the code input task, $64$\% reported being fine or okay with it, 
$8$\% wanted it removed, and $28$\% did not like it, though they recognized its security benefits. 

Most participants showed acceptance of both push notification and code input tasks, 
acknowledging their essential security benefits, despite finding them somewhat inconvenient.

\subsubsection{Open Ended Responses for Duo Push Notification Task}
From the open-ended responses regarding perceptions of the Duo push notification task, 
we identified four main themes: 1) Positive Sentiment, 2) Neutral Sentiment, 
3) Negative Sentiment, 4) Suggestions/Improvements.

 \begin{figure}[h]
        \centering
	{\includegraphics[width=\columnwidth]{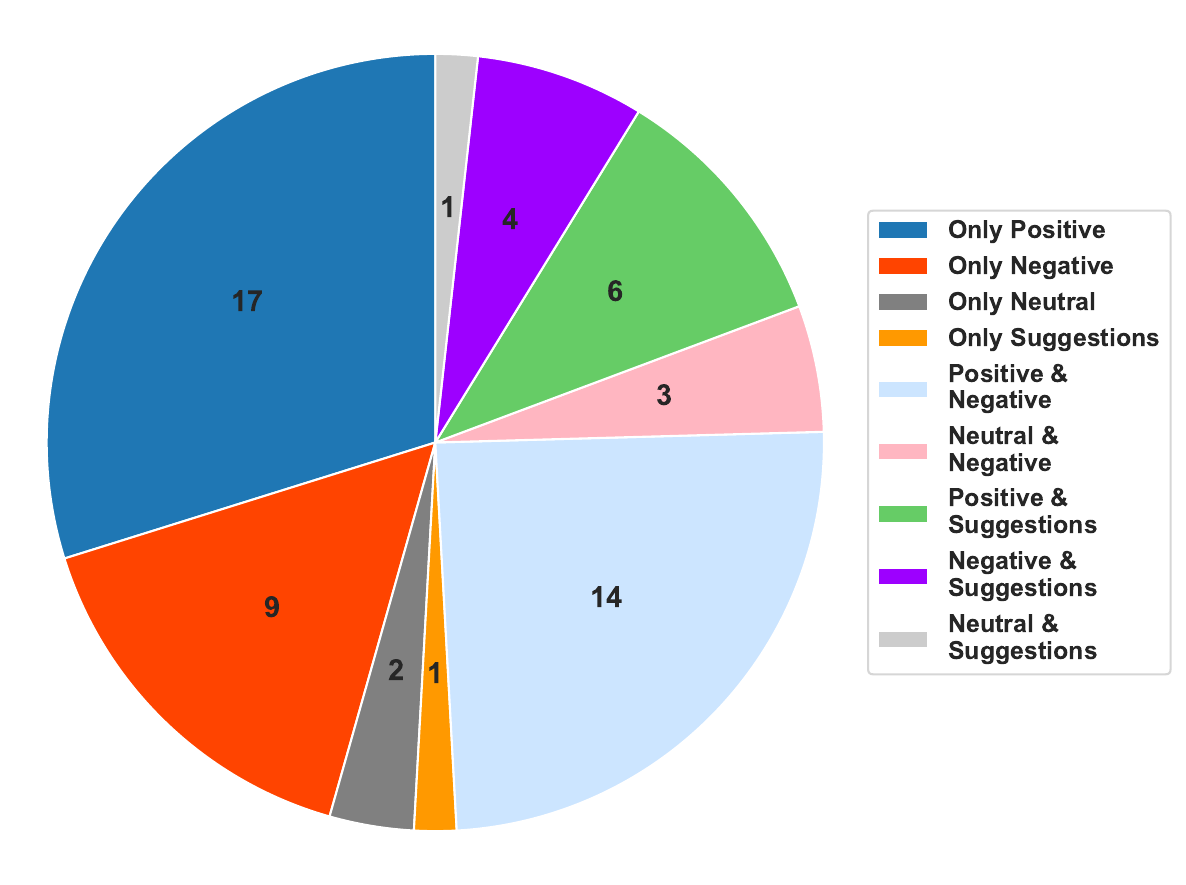}}
	
	\caption{Open Ended Responses: Perceptions of Duo Push Notification Task}
         
	\label{fig:open_ended_push} 
\end{figure}

Each response was categorized with one or more themes. Figure \ref{fig:open_ended_push} shows 
the division of responses according to associated themes. Among $57$ responses, the largest 
category consisted of responses with positive sentiment. Interestingly, the second largest category was responses 
expressing both positive and negative sentiments. $12$ responses included at least one suggestion 
for improvement.

One example of a response with both positive and negative sentiment is:
\begin{quote}
    ``It works well and works seamlessly with my Apple Watch. 
    It can be a pain when the service is poor."
\end{quote}
Another example response, with both neutral and negative sentiments:
\begin{quote}
    ``It's fine, it serves its purpose. It's annoying that I have to sign in multiple times per day, but it's easy enough to accept from the push notification."
\end{quote}
Suggestions for improvements included:
\begin{compactitem}
    \item Implementing a ``Remember Device'' feature
    \item Allowing acceptance directly from push notifications without opening the app
    \item Enabling fingerprint approval for Duo on desktop
    \item Making Duo functional without internet access
\end{compactitem}

\subsubsection{Open-Ended Responses for Duo Code Input Task}
\begin{figure}[h]
        \centering
	{\includegraphics[width=\columnwidth]{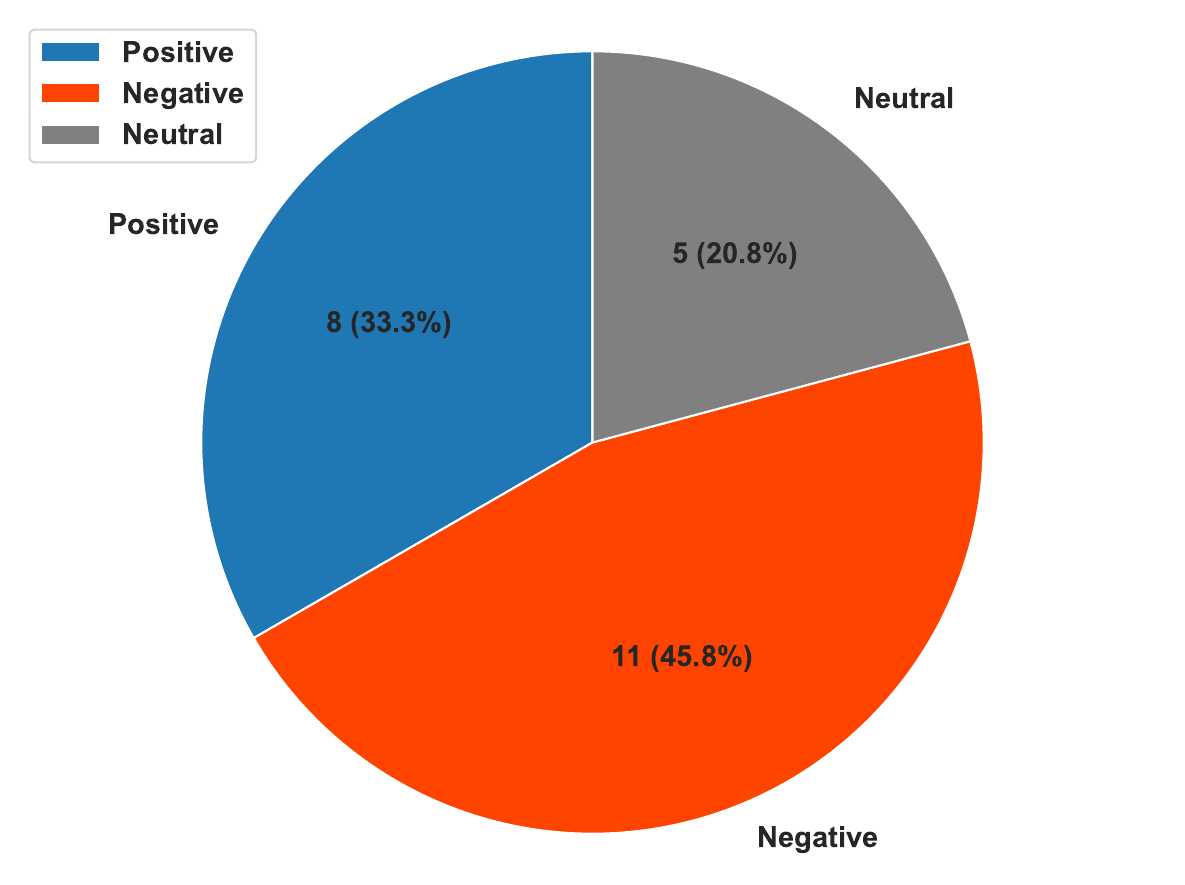}}
	\caption{Open Ended Responses: Perceptions of Duo Code Input Task}
         
	\label{fig:open_ended_code} 
\end{figure}

As before, responses fall into: 1) Positive Sentiment, 2) Neutral Sentiment, 3) Negative Sentiment.
Figure \ref{fig:open_ended_code} presents the response breakdown along these sentiments. The largest 
percentage of the responses conveys negative sentiment. Among those with positive sentiment, 
two participants indicated that, although they were okay with the task, they preferred 
push notification. As expected, the negative sentiment for the code input task 
is higher than that for push notification.

\subsection{Duo-induced Authentication Failure}
We analyzed authentication log files and participants' survey responses in order
to understand the percentage of authentication failures caused by Duo. We also 
investigated failure reasons via open-ended and multiple-choice questions.

\subsubsection{Failure Statistics: Log files} 
The message field in each log entry allows us to calculate the percentage of Duo
authentication failures. As mentioned earlier, log files contained $43,939$
entries for successful primary authentication and $1,913$ entries for 
Duo failure events. This denotes that $4.35$\% of authentication attempts failed 
due to Duo. Since this occurred after a user provided the correct userid and password, 
the failure rate appears to be significant.

To better understand the underlying causes of Duo authentication failures, we conducted a root cause analysis using the log files. The results are summarized in Table~\ref{tab: failure types}. The most prevalent issue, accounting for $84.89$\% of all failures, is the timeout of Duo Push Notification tasks. This indicates that users either did not respond to the push request in time, or that the notification was not successfully delivered.

The second most common failure type, responsible for $7.73$\% of cases, stems from user accounts being disabled or lacking the necessary permissions to access the Linux cluster. These failures were identified both during the authentication and pre-authentication phases.

The third major contributor, comprising $4.91$\% of failures, involves users who are either not enrolled in Duo or have misconfigured Duo settings, resulting in no usable authentication factors.

Other less frequent failure types include incorrect passcode entries and users actively denying the authentication request via the Duo app.

\begin{table}[h]
\caption{Duo Failure Types}
\label{tab: failure types}
\centering
 \begin{tabularx}{\linewidth}{p{7.3cm}p{2cm}}
\toprule  
  \textbf{Type} & \textbf{Count} \\
  \midrule
    Duo authentication was rejected - Login timed out. & 1624 \\  
    \midrule
    Duo preauth result was Your account is disabled and cannot access this application. Please contact your administrator. & 131 \\
    Duo authentication was rejected - Your account is disabled and cannot access this application. Please contact your administrator. & 17  \\ 
    \midrule
    User has no Duo factors usable with this configuration & 89\\
    Duo preauth result was Access is not allowed because you are not enrolled in Duo. Please contact your organization's IT help desk. & 5\\
     \midrule
    Duo authentication was rejected - Login request denied. & 42 \\
    \midrule
    Duo authentication was rejected - Incorrect passcode. Please try again. & 5  \\ 
\bottomrule
\end{tabularx}

\end{table}

\subsubsection{Failure Statistics: Survey Responses}
We asked survey participants whether they were ever unable to sign in 
because they could not complete Duo 2FA tasks. Out of $57$ participants, 
$25$ responded affirmatively to experiencing failures, which corresponds to $43.86$\%. This highlights that,
while Duo enhances security, it also causes occasional denial of service.

\subsubsection{Failure Reasons}
For participants who answered ``yes'' to experiencing Duo sign-in failures, we followed up 
by asking about exactly what happened, offering four options. Table \ref{tab: failure reasons} 
summarizes responses, with the majority indicating that they did not have access to their phone associated with their Duo account. This aligns closely with our log file analysis, which identified Duo Push Notification timeouts as the predominant failure type. Without access to their registered devices, users were unable to respond to the push notification before it expired, resulting in failed sign-in attempts.
\begin{table}[h]
\caption{Duo Failure Reasons}
\label{tab: failure reasons}
\centering
 \begin{tabularx}{\linewidth}{p{6cm}p{2cm}}
\toprule  
  \textbf{Option} & \textbf{\# Responses} \\
  \midrule
  Didn't have physical access to phone with Duo app & 16  \\
  Duo task did not appear on my phone & 5 \\  
  Duo task timed out before I could complete them & 1 \\  
  Others, please specify in the following answer & 3  \\ 
\bottomrule
\end{tabularx}

\end{table}
Those who chose the fourth option, 
``Other'', were given an open-ended follow-up question asking them to specify the failure 
reason. From those open-ended responses, we identified four additional issues:
\begin{compactitem}
    \item \textbf{Device Performance Issues}: Problems related to the phone's performance affecting app functionality (e.g., app unloading due to slow processing) -- reported by one participant.
    \item \textbf{Authentication Process Challenges}: Difficulties in completing the authentication process due to app unloading or other technical issues -- reported by one participant.
    \item \textbf{Notification Functionality Issues}: Problems with the notification system, such as the inability to approve/deny directly from notifications -- reported by one participant.
    \item \textbf{Connectivity Issues}: Problems related to a lack of access to 
    cellular data or Wi-Fi -- reported by two participants.
\end{compactitem}
Each participant provided multiple reasons for their sign-in failures. Notably, one 
reported experiencing all the issues mentioned in the previous multiple-choice question 
as well as additional connectivity problems. 

\subsection{Multiple Code Inputs in a Single Authentication Instance}
We asked $25$ participants who experienced the code input task whether they ever had to 
input the code multiple times. Ten participants answered yes. In addition, we asked them
participants to indicate the maximum number of codes they had to input in a single sign-in 
instance. Five participants selected $2$, three chose $3$, and two selected $6$ or more. This suggests 
that participants are sometimes trapped in a frustrating cycle of repeatedly having to enter 
$6$-digit codes.

\subsection{Perception of Duo Overhead}
\begin{figure}[htbp]
    \centering
    \begin{subfigure}{\columnwidth}
        \centering
        \includegraphics[width=\columnwidth]{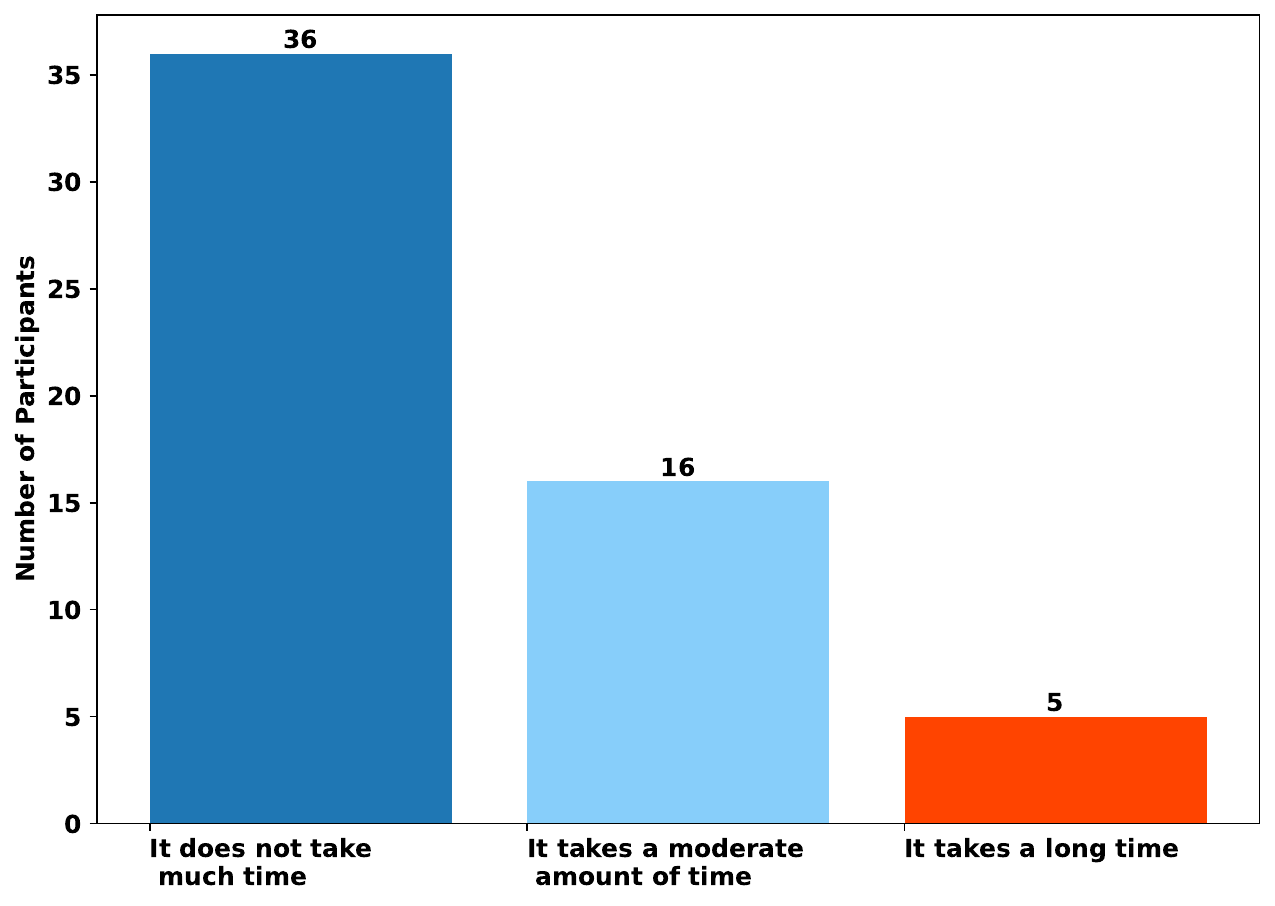}
        \caption{Duo Push Notification Task}
    \end{subfigure}
    \begin{subfigure}{\columnwidth}
        \centering
        \includegraphics[width=\columnwidth]{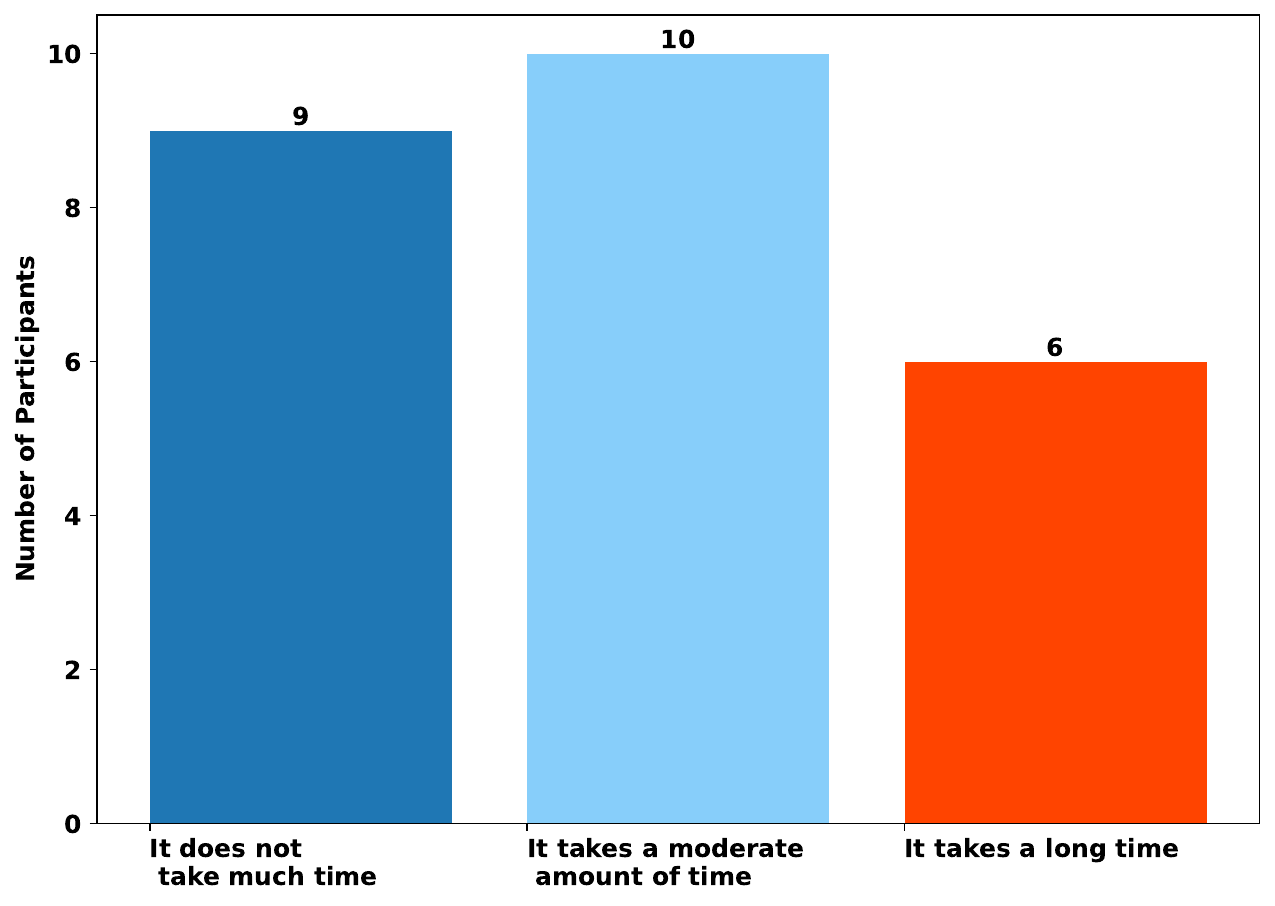}
        \caption{Duo Code Input Task}
    \end{subfigure}
    
    \caption{Participants' thoughts on Overheads of Duo}
    \label{fig:time_req}
\end{figure}

We surveyed participants regarding their opinions about the time to complete Duo push notification 
and code input tasks using multiple-choice questions. The goal was to investigate the participants' 
perception of the Duo overhead. Results are presented in Figure \ref{fig:time_req}. The majority of participants 
felt that Duo push notification tasks do not require much time. Conversely, responses about the Duo code 
input task were more evenly split between those who felt that (1) it does not take much time, or (2) it requires 
a moderate amount of time. Importantly, very few participants felt that either type of Duo task took a 
long time. Therefore, we can conclude that participants are generally not 
bothered by the overhead of a Duo task.

\subsection{Usage of 2FA systems outside of school}
We inquired whether participants used any other 2FA or MFA systems. Surprisingly, $38$ participants ($66.67$\%) 
responded ``Yes''. Therefore, many participants are likely accustomed to periodic MFA/2FA tasks, potentially 
making them more efficient and positively inclined towards completing Duo tasks, as compared to those who do not 
use 2FA outside of school.  

We also asked participants whether using Duo 2FA at school motivated them to adopt MFA/2FA for any of their non-school 
accounts. $40$ ($70.18$\%) replied ``No''. This suggests that -- while most users tolerate MFA/2FA when it is mandatory --
they are not inclined to adopt it voluntarily for security purposes. However, it is interesting that 
$\approx~30$\% were incentivized to voluntarily adopt 2FA for other accounts.

\subsection{Perception of Security}
\begin{figure}[h]
        \centering
	{\includegraphics[width=\columnwidth]{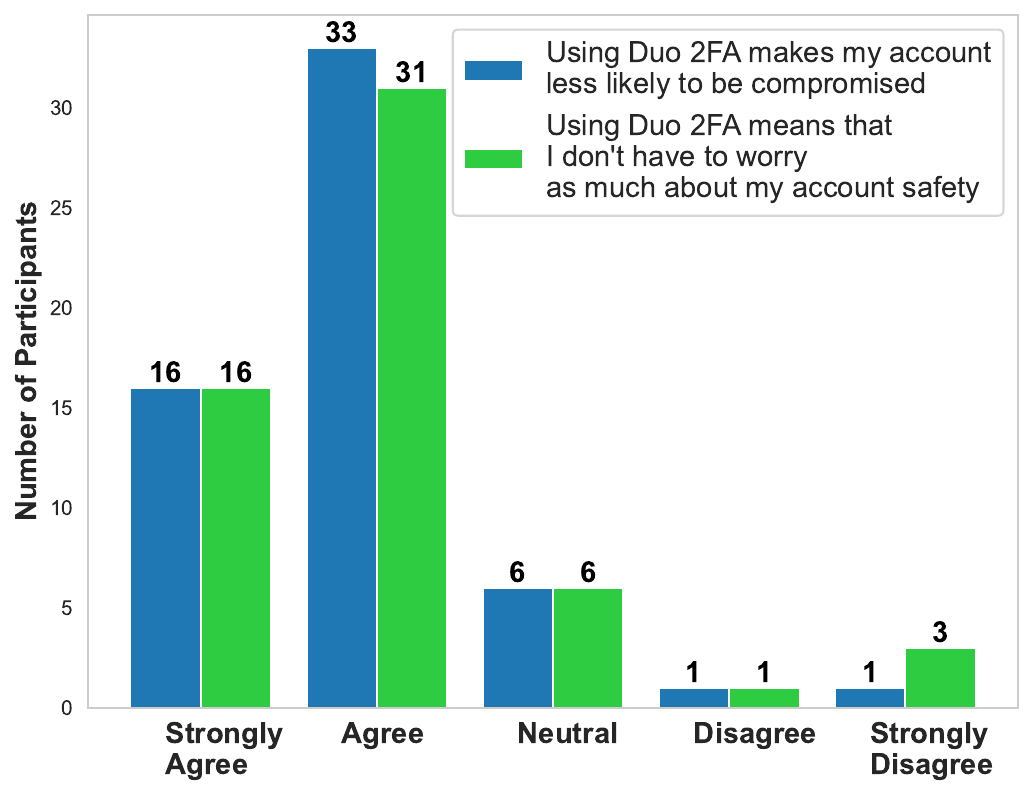}}
	\caption{Security Perceptions of Duo \\}
         
	\label{fig:secure_less_worry} 
\end{figure}

We asked participants two similar multiple-choice questions to assess how Duo affects their perception 
of account security.  Figure \ref{fig:secure_less_worry} provides a breakdown of responses. Participants responded 
very consistently: $77.19$\% selected either ``Agree'' or ``Strongly Agree'' for both questions. However, we found
that $3$ participants chose opposing options, e.g., agreeing in the first question and strongly disagreeing in the second. 
We believe that this is probably caused by inattention.
Despite this minor inconsistency, most participants believe that Duo enhances the security of their accounts.

\subsection{Perception of Usability}
\begin{figure}[h]
    \centering
    \includegraphics[width=\columnwidth]
    {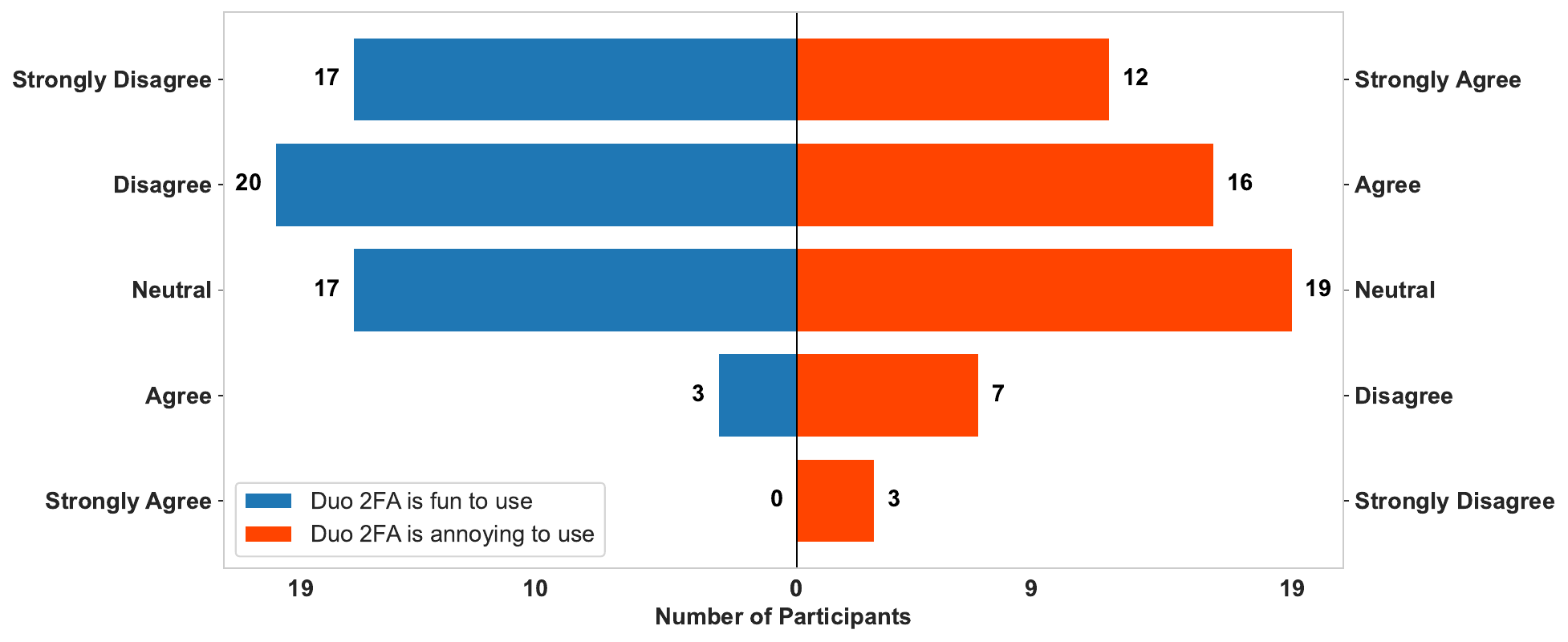}
    \caption{Duo 2FA is fun to use, Duo 2FA is annoying to use}
    \label{fig:fun_annoying}
\end{figure}

\begin{figure}[h]
    \centering
	{\includegraphics[width=\columnwidth]{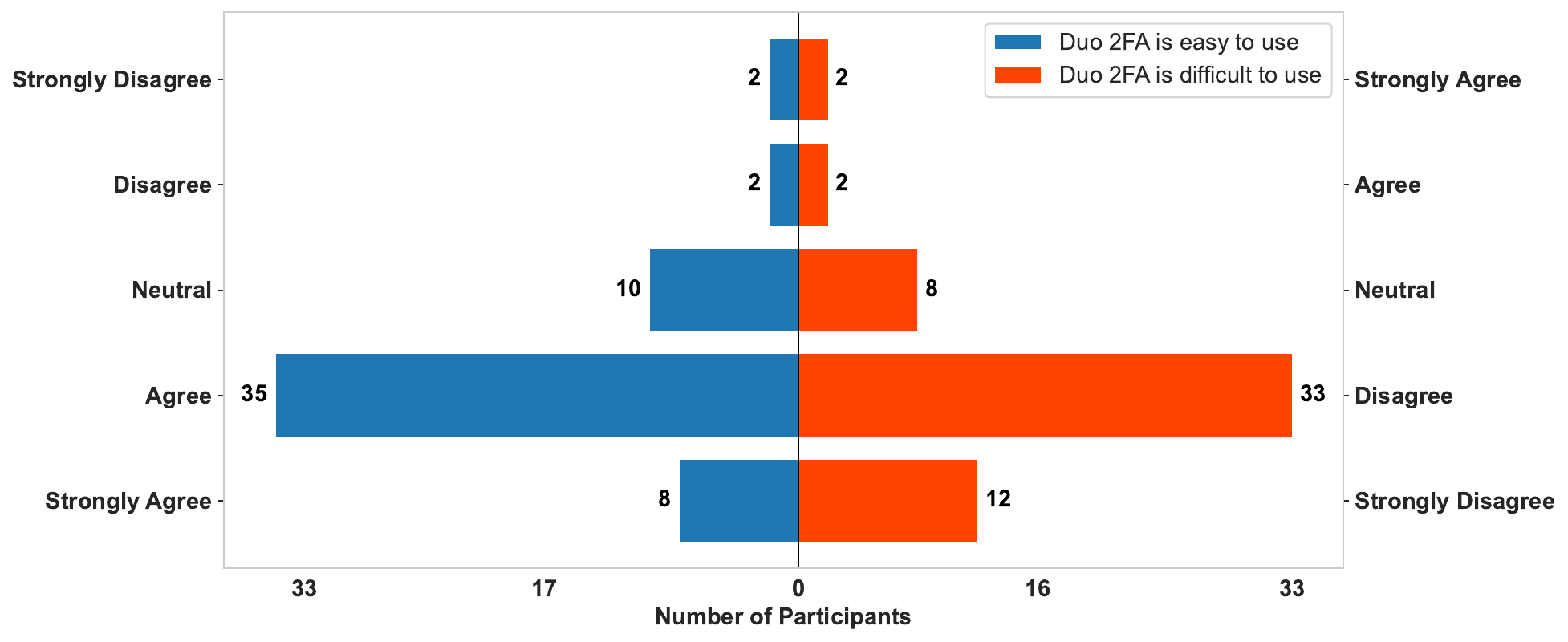}}
	\caption{Duo 2FA is easy to use, Duo 2FA is difficult to use}
	\label{fig:easy_difficult} 
\end{figure}

We asked participants to rate four statements to assess their perception of Duo usability. 
Results are presented in Figures \ref{fig:fun_annoying} and \ref{fig:easy_difficult}.

Most find Duo easy to use, yet annoying. Participants responded to these statements consistently, 
with the number of those who responded positively to Duo being ``easy'' closely matching the 
number of those who responded negatively to Duo being ``difficult''. Responses regarding 
Duo being ``fun'' versus ``annoying'' reflected a similar consistency. We also checked 
each participant's responses to the ``easy vs. difficult'' and ``fun vs. annoying'' 
statements for inconsistencies and found that all responses were consistent. This 
increases our confidence in the quality of the participants’ responses, suggesting 
that they did not just select options randomly.

\noindent \textbf{Usability Perception vs Experienced Push Overhead:}
We investigated whether Duo push overhead influenced participants' responses to these four statements. For example, we examined whether participants who selected ``Agree/Strongly Agree'' to the statement ``Duo is annoying to use'' experienced higher average Duo push overhead compared to those who selected ``Disagree/Strongly Disagree''. The results are presented in Table \ref{tab: usability overhead}.

\begin{table}[h]
\centering
\caption{Average Push Overhead (seconds) by Responses to Duo 2FA Usability Statements}
\begin{tabularx}{\linewidth}{p{1.8cm}p{1.1cm}p{0.9cm}p{0.8cm}p{0.7cm}p{1.1cm}}
\toprule
\textbf{Statement} & \textbf{Strongly Disagree} & \textbf{Disagree} & \textbf{Neutral} & \textbf{Agree} & \textbf{Strongly Agree} \\
\midrule
Duo 2FA is annoying to use & 7.98 & 8.34 & 7.36 & 8.038 & 7.33 \\
\midrule
Duo 2FA is difficult to use & 7.07 & 7.89 & 8.04 & 10.41 & 8.55 \\
\midrule
Duo 2FA is easy to use & 8.55 & 6.01 & 8.08 & 7.85 & 7.17 \\
\midrule
Duo 2FA is fun to use & 7.41 & 7.85 & 8.26 & 6.82 & NA     \\
\bottomrule
\end{tabularx}
\label{tab: usability overhead}

\end{table}

Our analysis revealed no discernible relationship between average push overhead and participants' responses. Furthermore, we found no statistically significant differences in push overhead values when participants were grouped by their responses to the usability statements. 

\subsection{Other Issues with Duo}
Finally, we included an optional open-ended question in the survey asking participants to describe any other issues or difficulties they had encountered with Duo. It is important to note that Duo is also used for second-factor authentication in the single sign-on system of the university where this study was conducted. Consequently, some issues reported by participants involved challenges with the central system. We received ten responses, and the following themes were identified from these responses:

\begin{compactenum}
    \item \textbf{Ease of Use}: Mentions of Duo being easy to use.
    
    \item \textbf{Frequency of Use}: Annoyance with having to use Duo multiple times a day.

    \item \textbf{Single Sign-On (SSO) Issues}: Frustrations with the university's single sign-on 
    not working as expected across systems.

    \item \textbf{Manual Login Issues}: Problems with having to manually enter login information in apps.

    \item \textbf{Device Change Issues}: Difficulties encountered when changing phones or phone numbers.

    \item \textbf{Phone Dependency}: Inability to log in without the phone.

    \item \textbf{Notification Issues}: Problems with receiving push notifications, including delays or failures.

    \item \textbf{App Compatibility}: Mention of app availability on devices like the Galaxy Watch.

    \item \textbf{Network Issues}: Problems related to network issues affecting Duo's functionality.

    \item \textbf{Overall Satisfaction}: General satisfaction or lack of difficulties with using Duo.
\end{compactenum}
Three responses mentioned notification issues, and two -- highlighted device change issues (i.e., difficulties 
in changing a registered phone number or device). Other themes were reported by one response each.

\subsection{SUS Score}
We used SUS to quantitatively assess Duo usability. The SUS score computed
from the survey is $70$, indicating a ``Good'' usability level. This score aligns 
with participants' perceptions that Duo tasks are easy and quick to complete. 
However, the perception of Duo being annoying likely contributes to it not 
achieving an ``Excellent'' usability level, despite the tasks being 
straightforward and time-efficient. We also examined the potential correlation between participants' experienced overheads and their reported SUS scores. No significant relationship was observed between these two measures.  
\section{Comparison with Related Work}\label{sec:related_work}
Our research focuses on two primary areas: (1) quantifying the overhead associated with Duo,
and (2) examining the usability and user perception of Duo. Consequently, we compare our study 
with prior research within these two domains. Table \ref{tab:related_work} presents an overview of the comparison.

\begin{table*}[h]
\caption{Comparison with Related Work}
\label{tab:related_work}
\centering\begin{tabularx}{\textwidth}{p{0.6cm} p{1.5cm} p{1.5cm} p{2cm} p{2cm} p{2.5cm} p{4.8cm}}
\toprule
&\textbf{\#Participants} & \textbf{Year} & \textbf{Push Overhead (seconds)} & \textbf{Failure Rate (log, survey)} & \textbf{Push SUS Score} & \textbf{User Perception} \\
\midrule
\textbf{this} & \textbf{2559} & \textbf{2024-2025} & \textbf{7.82} & \textbf{4.35\%, 43.86\%} & \textbf{70} & \textbf{easy, annoying, increases security} \\
\midrule
\cite{reynolds2020empirical} & $>$77K & 2018 & NA & $>5$\%, NA & NA & NA\\
\midrule
\cite{acemyan20182fa} & 27 & 2017 & 58 & 5\%, NA & 85 & NA\\
\midrule
\cite{reese2019usability} & 72 & 2018 & 11.8 & NA & 81 & increases security\\ 
\midrule
\cite{colnago2018s} & 2047 & 2017 & NA & NA & NA & easy, annoying, increases security\\
\midrule
\cite{dutson2019don} & 4275 & 2018 & NA & NA, 50\% & NA & easy, increases security\\
\midrule
\cite{abbott2020mandatory} & 565 & 2016-2018 & NA & 10\%, NA & NA & NA \\ 
\midrule
\cite{arnold2022emotional} & 465 & 2021 & NA & NA & NA & increases stress, anxiety, frustration\\
\midrule
\cite{das2019mfa} & NA, 12500 user reviews & 2019 & NA & NA & NA & setup difficulty, backup and migration issue, forced to use, demand for device compatibility \\
\midrule
\cite{marky2022nah} & 42 & 2021 & NA & NA & NA & requirements: excellent usability, location-independence, functional without internet, independence from energy source, secure interaction\\
\bottomrule
\end{tabularx}
\vspace{-0.2in}
\end{table*}

\subsection{MFA/2FA Overhead}
Prior studies reported 2FA overheads ranging from $11$ seconds to $58$ seconds \cite{acemyan20182fa, reese2019usability, kruzikova2024two}, notably higher than our calculated overhead of $7.82$ seconds. Acemyan et al. \cite{acemyan20182fa} reported a $58$ seconds average for Gmail login using Google Prompt and SMS, though this included the entire login sequence. Reese et al. \cite{reese2019usability} utilized a simulated banking environment to find a median push overhead of $11.8$ seconds. Recent work by Kruzikova et al. \cite{kruzikova2024two} examined hardware-based 2FA (NFC and payment cards), finding overheads between 11 seconds and $35$ seconds; they concluded that higher ``time satisfaction'' directly correlates with a higher perceived security. 

Reynolds et al. \cite{reynolds2020empirical} used university log data to mathematically estimate annual time-costs rather than direct measurement, yet their findings align with ours regarding technical friction, noting a failure rate exceeding $5$\%. Our significantly lower overhead of $7.82$ seconds likely reflects a combination of improved UI designs and a more technologically adept user base compared to studies conducted $5$--$7$ years ago.

\subsection{MFA/2FA Usability and Perception}
While prior studies of push-based 2FA report ``excellent'' usability with SUS scores between $81$ and $85$ \cite{acemyan20182fa, reese2019usability}, our Duo-specific SUS score of $70$ indicates a ``good'' but lower usability level. This discrepancy may be explained by the mandatory nature of Duo in university settings. Colnago et al. \cite{colnago2018s} and Dutson et al. \cite{dutson2019don} found that while users find Duo ``easy'', they simultaneously find it ``annoying''. This annoyance is exacerbated during mandatory rollouts; Abbott et al. \cite{abbott2020mandatory} observed that user acceptance degrades significantly when 2FA moves from an optional or selective phase to a universal requirement.

Social and emotional factors also play a role. Weidman et al. \cite{weidman2017like} highlighted the tension of using personal devices for work-related 2FA, which many users view as an infringement on personal life. Arnold et al. \cite{arnold2022emotional} found that 2FA can induce stress and anxiety, particularly when technical failures hinder time-sensitive tasks. Large-scale sentiment analysis by Das et al. \cite{das2019mfa} supports this, revealing that Duo holds a middle-tier sentiment score ($0.349$) compared to other authenticators, with users frequently citing backup and device compatibility issues. 

Finally, our findings on connectivity concerns mirror Marky et al. \cite{marky2022nah}, who identified location-independence and offline functionality as primary user requirements. Other research continues to explore the lack of uniform design patterns across 2FA journeys \cite{ghorbani2023systematic} and the potential for lifecycle management tools to improve the setup and removal process \cite{smith2023if, das2019evaluating, das2020mfa, wahab2023user, jensen2021multi, klivan2023we}.
\section{Discussion}\label{sec:discussion}

\subsection{Limitations}
While this study draws on extensive real world Duo usage and combines fine grained log analysis with user survey data, 
the following limitations should be taken into account when interpreting our findings.

\noindent \textbf{Demographic Bias.} The sample is strongly skewed toward technically proficient users: over $98\%$ 
were students (mostly majoring in Computer Science or related disciplines) who use Duo frequently. They may be more 
able and willing to work around Duo friction than older adults, non‑technical types, or those with lower digital literacy.

\noindent \textbf{Survey Response Bias.} The data sources are imbalanced: logs cover $2,500+$ users, but only $57$ completed 
the survey which corresponds to $~2.2\%$ response rate. This likely introduces volunteer bias (e.g., stronger opinions, 
higher security awareness, or more extreme experiences), so perceived overhead, annoyance, security benefit, and
SUS should be treated as indicative rather than representative.

\noindent \textbf{Limited Insight into Overhead Sources.} Retrospective logs measure only aggregate push overhead, 
not its components. The $~7.82$-second average delay combines user behavior, device performance, and network/server 
latency. Logs lack device/OS information, network conditions, and whether approval happened via notification 
or by opening the app. Thus, we cannot attribute delay to inattention, notification issues, UI design, or the infrastructure.

\noindent \textbf{Lack of Accessibility Evaluation.} Accessibility and inclusivity were not assessed. We did not recruit/identify 
users with visual, motor, or cognitive impairments or observe assistive technology use. So we cannot conclude how usable Duo is 
for people with disabilities.

\subsection{Recommendations for Improving Duo MFA Usability}
Duo MFA achieved a ``Good'' usability level (SUS=70) and low average authentication overhead ($7.82$ seconds). 
However, log data and user feedback indicate recurring sources of friction. The following suggested
improvements are intended to reduce user burden while preserving Duo’s security guarantees.
 
\noindent \textbf{Reducing Prompt Frequency.} 
Repeated daily MFA prompts were a common complaint. ``Remember device'' and risk‑based policies can be enabled to 
suppress prompts for trusted devices and low‑risk contexts (e.g., same browser, device, and network for $8–24$ hours). 
Although this capability is supported by Duo, it was not enabled in the studied deployment. Appropriate ``remember windows''
should be configured based on the institutional risk tolerance (e.g., longer intervals for staff desktops and shorter ones 
for shared lab machines), and the conditions under which a new prompt is required should be clearly communicated to 
maintain user trust.
 
\noindent \textbf{Improving Push Reliability and Timeout Recovery.} Push notification timeouts constituted the largest category of Duo‑related authentication failures. Push delivery robustness can be increased through retries across multiple channels and through clear on‑screen status indicators (e.g., request sent, waiting for approval, push failed). When a push is likely to time out, fallback options (e.g., passcode entry or use of a secondary device) should be offered proactively rather than requiring a full restart of the login sequence.
 
\noindent \textbf{Supporting Offline and Low‑Connectivity Scenarios.} Delays and lockouts were reported in low‑connectivity contexts. Offline options should be made more prominent during setup and within the authentication interface, including guidance for generating passcodes in advance and storing them safely for travel or field work. When network problems are detected, targeted guidance (e.g., recommending passcode use instead of waiting for a push) should be presented to reduce avoidable failures.
 
\noindent \textbf{Strengthening Fallbacks When the Phone Is Unavailable.} Lack of access to the registered phone was the dominant user‑reported reason for being unable to complete Duo tasks, aligning with failure modes observed in the logs. Enrollment of at least one additional factor (e.g., a backup device, hardware token, or landline) should be encouraged, and periodic reminders to verify backup validity should be incorporated. Self‑service recovery flows based on pre‑registered secondary factors should be provided to reduce lockouts and administrative burden while maintaining security.
 
\noindent \textbf{Lowering Interaction Cost at Approval Time.} Many participants reported friction when phone unlock and app‑opening were required for approval, particularly when in‑notification “Approve/Deny” buttons did not work. Reliable in‑notification approval actions would reduce interaction cost and better align with user expectations. Biometric confirmation on mobile or desktop platforms (e.g., Touch ID or Windows Hello) can also be integrated to streamline approvals; these features are configurable but were not utilized in the studied deployment.
 
\noindent \textbf{Clarifying Passcode‑Based Authentication Flows.} Multiple passcode entries within a single session were reported by some participants. Interfaces should distinguish clearly between expired and invalid codes, and passcode prompts should be limited to explicit fallback scenarios rather than repeated without explanation. Contextual feedback should be provided to indicate next steps (e.g., generating a new code or carefully re‑entering the current code), and recovery behavior should be streamlined when a code fails.
 
These recommendations translate empirical failure patterns and user feedback into design and deployment changes that can improve everyday usability without weakening Duo’s security guarantees.
\section{Conclusion}\label{sec:conclusion}
This paper reports on a comprehensive usability analysis of Duo 2FA, focusing specifically on the 
overhead of Duo push notification tasks, the factors influencing this overhead, authentication 
failures associated with Duo, the SUS score, and user perceptions of the platform. Our research used
authentication log data from UCI, involving $2,559$ unique users over a nine-month period, from August 
2024 to April 2025. Also, we administered a survey to $57$ randomly selected participants to gather 
qualitative insights. Our findings indicate that the average overhead of a Duo push notification task is 
$7.82$ seconds, influenced by factors such as the time of day, participants' academic major, and educational levels. 
The system achieved a SUS score of $70$, reflecting a ``Good'' level of usability. Notably, $43.86\%$ of 
participants reported experiencing at least one authentication failure due to an incomplete Duo task. 
While users acknowledged that Duo tasks were generally easy to complete, they also described the tasks 
as annoying. Nevertheless, Duo was found to enhance users' sense of security regarding their accounts. 
Participants also identified common issues, such as needing to open the Duo app when a push notification 
fails and experiencing difficulties completing Duo tasks due to occasional lack of Internet access. They 
proposed several improvements, including a ``Remember Device'' feature.

\noindent {\bf Acknowledgements:} 
We thank USEC’26 reviewers for their constructive feedback. This work was supported 
in part by funding from the NSF Award SATC-1956393.
\bibliographystyle{unsrt}
\bibliography{reference}
\appendix

\section*{Survey Questionnaire}
\begin{compactitem}
    \item 2FA: Two-Factor Authentication
    \item MFA: Multi-Factor Authentication
\end{compactitem}


\subsection*{Demographics}
\begin{compactenum}
    \item What is your age?
    \item What is the highest level of education you attained?
    \begin{compactitem}
        \item Below High School
        \item High School
        \item Associate Degree
        \item Bachelor Degree
        \item Master Degree
        \item Doctorate (PhD)
    \end{compactitem}
    \item What is your gender?
    \begin{compactitem}
        \item Male
        \item Female
        \item Non-binary
        \item Transgender
        \item Prefer not to say
    \end{compactitem}
\end{compactenum}

\subsection*{General Frequency and Opinion}
\begin{compactenum}
    \item How frequently do you interact with Duo to access UCI services?
    \begin{compactitem}
        \item a few times per month or less
        \item a few times per week
        \item 1 to 2 times daily 
        \item 3 to 5 times daily
    \end{compactitem}
    \item What do you think of Duo 2FA system?
\end{compactenum}

\subsection*{Detailed Experiences}
\begin{compactenum}
    \item What is your opinion about the time required to complete Duo 2FA tasks?
    \begin{compactitem}
        \item It does not take much time
        \item It takes a moderate amount of time
        \item It takes a long time
    \end{compactitem}
    \item Have you used any other 2FA or MFA system before encountering Duo 2FA at UCI?
    \begin{compactitem}
        \item Yes
        \item No
    \end{compactitem}
    \item Have you been motivated by using Duo at UCI to adopt MFA/2FA for any of your non-UCI accounts?
    \begin{compactitem}
        \item Yes
        \item No
    \end{compactitem}
    \item What is your opinion about completing Duo tasks as part of UCI access?
    \begin{compactitem}
        \item I am fine with it
        \item I am okay with it but find it somewhat inconvenient
        \item I do not like it but understand that it's needed for security purposes
        \item I do not like it and would prefer it to be removed
    \end{compactitem}
\end{compactenum}
\subsection*{Specific Incidents}
\begin{compactenum}
    \item Were you ever unable to login because it wasn't possible to complete Duo 2FA tasks?
    \begin{compactitem}
        \item Yes
        \item No
    \end{compactitem}
    \item If yes, which of the following happened?
    \begin{compactitem}
        \item I did not have physical access to the phone registered with Duo
        \item Duo task did not appear on my phone
        \item Duo task timed out before I could complete them
        \item Other, please specify
    \end{compactitem}
\end{compactenum}

\subsection*{Opinion and Experience Details}
Please rate the following statements about Duo 2FA:

\begin{compactenum}
    \item Using Duo 2FA makes my account less likely to be compromised. 
    \begin{compactitem}
        \item Strongly Disagree
        \item Disagree
        \item Neutral
        \item Agree
        \item Strongly Agree
    \end{compactitem}
    
    \item Using Duo 2FA means that I don't have to worry as much about my account safety. 
    \begin{compactitem}
        \item Strongly Disagree
        \item Disagree
        \item Neutral
        \item Agree
        \item Strongly Agree
    \end{compactitem}
    
    \item Duo 2FA is fun to use.
    \begin{compactitem}
        \item Strongly Disagree
        \item Disagree
        \item Neutral
        \item Agree
        \item Strongly Agree
    \end{compactitem}
    
    \item Duo 2FA is easy to use.
    \begin{compactitem}
        \item Strongly Disagree
        \item Disagree
        \item Neutral
        \item Agree
        \item Strongly Agree
    \end{compactitem}
    
    \item I think that Duo 2FA is difficult to use.
    \begin{compactitem}
        \item Strongly Disagree
        \item Disagree
        \item Neutral
        \item Agree
        \item Strongly Agree
    \end{compactitem}
    
    \item I think that Duo 2FA is annoying to use.
    \begin{compactitem}
        \item Strongly Disagree
        \item Disagree
        \item Neutral
        \item Agree
        \item Strongly Agree
    \end{compactitem}
\end{compactenum}
\subsection*{System Usability Scale}
In regard to Duo 2FA, denoted as 'system' below, please score the following 10 items:

\begin{compactenum}
    \item I think that I would like to use this system frequently:
    \begin{compactenum}
        \item Strongly Disagree
        \item Disagree
        \item Neutral
        \item Agree
        \item Strongly Agree
    \end{compactenum}

    \item I found the system unnecessarily complex:
    \begin{compactenum}
        \item Strongly Disagree
        \item Disagree
        \item Neutral
        \item Agree
        \item Strongly Agree
    \end{compactenum}

    \item I thought the system was easy to use:
    \begin{compactenum}
        \item Strongly Disagree
        \item Disagree
        \item Neutral
        \item Agree
        \item Strongly Agree
    \end{compactenum}

    \item I think that I would need the support of a technical person to be able to use this system:
    \begin{compactenum}
        \item Strongly Disagree
        \item Disagree
        \item Neutral
        \item Agree
        \item Strongly Agree
    \end{compactenum}

    \item I found that the various functions in this system were well integrated:
    \begin{compactenum}
        \item Strongly Disagree
        \item Disagree
        \item Neutral
        \item Agree
        \item Strongly Agree
    \end{compactenum}

    \item I thought there was too much inconsistency in this system:
    \begin{compactenum}
        \item Strongly Disagree
        \item Disagree
        \item Neutral
        \item Agree
        \item Strongly Agree
    \end{compactenum}

    \item I would imagine that most people would learn to use this system very quickly:
    \begin{compactenum}
        \item Strongly Disagree
        \item Disagree
        \item Neutral
        \item Agree
        \item Strongly Agree
    \end{compactenum}

    \item I found the system very cumbersome to use:
    \begin{compactenum}
        \item Strongly Disagree
        \item Disagree
        \item Neutral
        \item Agree
        \item Strongly Agree
    \end{compactenum}

    \item I felt very confident using the system:
    \begin{compactenum}
        \item Strongly Disagree
        \item Disagree
        \item Neutral
        \item Agree
        \item Strongly Agree
    \end{compactenum}

    \item I needed to learn a lot of things before I could get going with this system:
    \begin{compactenum}
        \item Strongly Disagree
        \item Disagree
        \item Neutral
        \item Agree
        \item Strongly Agree
    \end{compactenum}
\end{compactenum}
\end{document}